\documentclass[fleqn,usenatbib,referee]{mnras}

\usepackage[T1]{fontenc}

\DeclareRobustCommand{\VAN}[3]{#2}
\let\VANthebibliography\thebibliography
\def\thebibliography{\DeclareRobustCommand{\VAN}[3]{##3}\VANthebibliography}

\usepackage{graphicx}

\usepackage{amsmath}    
\usepackage{amssymb}    

\title[Refining the OJ 287 2022 impact flare arrival epoch]{Refining the OJ 287 2022 impact flare arrival epoch}

\author[M. Valtonen et al.]{Mauri J. Valtonen,$^{1,2}$\thanks{E-mail: mvaltonen2001@yahoo.com (MJV)} 
Staszek Zola,$^3$ A. Gopakumar,$^4$
Anne L\"ahteenm\"aki,$^{5,6}$
\newauthor{Merja Tornikoski,$^5$ Lankeswar Dey,$^4$ Alok C. Gupta,$^{7,20}$ Tapio Pursimo,$^8$ Emil Knudstrup,$^8$}
\newauthor{Jose L. Gomez,$^9$ Rene Hudec,$^{10,11}$ Martin Jel\'{\i}nek,$^{11}$ Jan \v{S}trobl,$^{11}$ Andrei V. Berdyugin,$^2$}
\newauthor{Stefano Ciprini,$^{12,13}$ Daniel E. Reichart,$^{14}$ Vladimir V. Kouprianov,$^{14}$}
\newauthor{Katsura Matsumoto,$^{15}$ Marek Drozdz,$^{16}$ Markus Mugrauer,$^{17}$ Alberto Sadun,$^{18}$}
\newauthor{Michal Zejmo,$^{19}$ Aimo Sillanp\"a\"a,$^2$ Harry J. Lehto,$^2$ Kari Nilsson,$^1$}
\newauthor{Ryo Imazawa,$^{21}$  and Makoto Uemura $^{22}$}
\\
$^1$ FINCA, University of Turku, Turku, Finland\\
$^2$ Tuorla Observatory, Department of Physics and Astronomy, University of Turku, Turku, Finland\\
$^3$ Astronomical Observatory, Jagiellonian University, ul. Orla 171, 30-244 Krakow, Poland\\
$^4$ Department of Astronomy and Astrophysics, Tata Institute of Fundamental Research, Mumbai, India\\
$^5$ Aalto University, Mets\"ahovi Radio Observatory, Mets\"ahovintie 114, 02540 Kylm\"al\"a, Finland\\
$^6$ Aalto University, Department of Electronics and Nanoengineering, P.O. Box 15500, FI-00076 AALTO, Finland\\
$^7$ Aryabhatta Research Institute of Observational Sciences (ARIES), Manora Park, Nainital 263001, India\\
$^8$ Nordic Optical Telescope, Apartado 474, E-38700 Santa Cruz de La Palma, Spain\\
$^9$ Instituto de Astrofisica de Andalucia - CSIC, Glorieta de la Astronomia s/n, 18008 Granada, Spain\\
$^{10}$ Czech Technical University, Faculty of Electrical Engineering, Prague, Czech Republic\\
$^{11}$ Astronomical Institute (ASU CAS), Ond\v{r}ejov, Czech Republic\\
$^{12}$ Instituto Nazionale di Fisica Nucleare (INFN) Sezione di Roma Tor Vergata, Via della Ricerca Scientifica 1,     \\00133, Roma, Italy\\
$^{13}$ ASI Space Science Data Center (SSDC), Via del Politecnico, 00133, Roma, Italy\\
$^{14}$ University of North Carolina at Chapel Hill, Chapel Hill, North Carolina, NC 27599, USA\\
$^{15}$ Astronomical Institute, Osaka Kyoiku University, 4-698 Asahigaoka, Kashiwara, Osaka, 582-8582, Japan\\
$^{16}$ Mt. Suhora Observatory, Pedagogical University, ul. Podchorazych 2, 30-084 Krakow, Poland\\
$^{17}$ Astrophysikalisches Institut und Universitäts-Sternwarte, Schillergässchen 2, D-07745 Jena, Germany\\
$^{18}$ Department of Physics, University of Colorado, Denver, CO 80217, USA\\
$^{19}$ Kepler Institute of Astronomy, University of Zielona Gora, Lubuska 2, 65-265 Zielona Gora, Poland\\
$^{20}$ Key Laboratory for Research in Galaxies and Cosmology, Shanghai Astronomical Observatory,\\ Chinese Academy of Sciences, Shanghai 200030, China\\
$^{21}$ Department of Physics, Graduate School of Advanced Science and Engineering, Hiroshima University,\\ 1-3-1 Kagamiyama, Higashi-Hiroshima, Hiroshima 739-8526, Japan\\
$^{22}$ Hiroshima Astrophysical Science Center, Hiroshima University, 1-3-1 Kagamiyama, Higashi-Hiroshima,\\ Hiroshima 739-8526, Japan
}

\date{Accepted ... Received ... in original form ...}

\pubyear{2022}

\begin{document}
\label{firstpage}
\pagerange{\pageref{firstpage}--\pageref{lastpage}}
\maketitle

\begin{abstract}
The bright blazar OJ~287 routinely parades high brightness bremsstrahlung flares, which are explained as being a result of a secondary supermassive black hole (SMBH) impacting the accretion disc of a more massive primary SMBH in a binary system. The accretion disc is not rigid but rather bends in a calculable way due to the tidal influence of the secondary. Below we refer to this phenomenon as a variable disc level. We begin by showing that these flares occur at times predicted by a simple analytical formula, based on general relativity inspired modified 
Kepler equation, which explains impact flares since 1888.  
 The 2022 impact flare, namely flare number 26, is rather peculiar as it breaks the typical pattern of two impact flares per 12-year cycle.  This is the third bremsstrahlung flare of the current cycle that follows the already observed 2015 and 2019 impact flares from OJ~287.
 It turns out that the arrival epoch of flare number 26 
 is sensitive to the level of primary SMBH's accretion disc relative to its mean level 
 in our model.
We incorporate these tidally induced changes in the level of the accretion disc to infer that the 
thermal flare should have occurred during July-August 2022, when it was not possible to observe it from the Earth.
Thereafter, we explore possible 
observational evidence for certain pre-flare activity by 
employing spectral and polarimetric data from our campaigns in 2004/05 and 2021/22. 
We point out theoretical and observational implications of two observed mini-flares during 
January-February 2022.

 \end{abstract}

\begin{keywords}{

BL Lacertae objects: individual: OJ~287 -- quasars: supermassive black holes -- accretion, accretion discs -- gravitational waves -- galaxies: jets
} 
\end{keywords}

\section{Introduction} \label{sec:intro}

Supermassive black hole (SMBH) binary systems are expected in the standard cosmological scenario as most massive galaxies contain a SMBH at their center and binaries should form by the merger of these galaxies \citep{BBR80,val89,mik92,val96b,qui96,mil01,vol03,kz2016,bur18}. Electromagnetic observations suggest the existence of more than a dozen SMBH binary candidates in active galactic nuclei \citep{ Koss23,charisi2016,Graham2015,Zhu2020,Bon2016,Liu2014, lai99,kau17}.

In contrast, there are only a few candidates that are compact enough to emit nano-hertz gravitational waves  \citep{val21,Iguchi2010, 
ONeill22}.

However, detailed theoretical investigations and observational campaigns make OJ~287, a BL Lacertae object at a redshift of 0.306 \citep{sit85,nil10}, a very special GW induced inspiraling 
SMBH binary candidate \citep{val21}. Interestingly, the binary nature of the OJ~287 central engine was recognised by one of us (Aimo Sillanp\"a\"a) already back in 1982, while constructing historical light curves for the quasars in the Tuorla - Mets\"ahovi variability survey, which had begun two years earlier \citep{kid07}. This inference was based on the observational evidence for major flares around 1911, 1923, 1935, 1947, 1959, and 1971 in the historical light curve of OJ~287. From this sequence it was easily extrapolated that OJ~287 should display a major outburst in 1983. The blazar monitoring community was alerted, resulting in a successful observational campaign of OJ~287. Indeed, one of biggest flares ever observed in OJ~287 occurred at the beginning of 1983 \citep{sil85,smi85}.
Following this success, further flares were predicted by \cite{sil88}, the next one in the autumn of 1994. It was indeed verified by the second campaign called OJ-94 \citep{sil96a}.

It was recognised soon after that these flares in OJ~287 were not exactly periodic, and that the systematics of the past flares are better understood if the flares come in pairs separated by $\sim 1-2$ years \citep{val96,LV96}. This led to the proposal of a new SMBH binary  central engine model for OJ~287, where the secondary SMBH orbits the more massive primary SMBH in a relativistic eccentric orbit with a redshifted orbital period of $\sim 12$ years. The orbital plane is inclined with respect to the accretion disk of the primary at a large angle, which leads to the secondary SMBH impacts with the accretion disc of the primary twice every orbit. These impacts lead to the pairs of flares in OJ~287. The next campaign, carried out by the OJ-94 group, verified the flare on October 1995, the second one of the pair. Interestingly, it came within the narrow two-week time window of the prediction \citep{val96,sil96b}.

 Subsequently, a number of investigations were pursued to improve astrophysical, observational, and theoretical aspects of the SMBH binary central engine description for OJ~287  \citep{Ram07,val10,Hudec2013,Pursimo2000,val06,val06a,val08,val11b,val16,Laine20,dey19}. These efforts allowed us to obtain the following values for OJ~287's SMBH binary system:
primary mass $m_1 = 18.35\pm0.05 \times 10^9 M_{\odot}$, secondary mass $m_2 = 150\pm10 \times 10^6 M_{\odot}$, primary Kerr parameter $\chi_1 = 0.38\pm0.05$, orbital eccentricity $e = 0.657\pm0.003$, and orbital period (redshifted) $P = 12.06\pm0.01$ years \citep{val10,Dey18}. These are among the nine parameters of a unique mathematical solution that can be extracted from the observed 
timing of 10 optical flares.

Let us emphasize that an acceptable solution exists 
if and only if each of these  ten flares comes within a narrow time window, whose width is specified in  \cite{Dey18}.

Further, it turns out that the up-to-date SMBH binary orbital description is consistent with additional seven flare epochs
which implies that the model is strongly over-determined.
\citep{dey19}.

The resulting impact flare epoch sequence, extracted from \cite{Dey18}, reads:  
1886.62 (1), 1896.67 (2), 1898.61 (3), 1906.20 (4), 1910.59 (5), 1912.98 (6), 1922.53 (7), 1923.73 (8), 1934.34 (9), 1935.40 (10), 1945.82 (11), 1947.28 (12), 1957.08 (13), 1959.21 (14), 1964.23 (15), 1971.13 (16), 1972.93 (17), 1982.96 (18), 1984.12 (19), 1994.59 (20), 1995.84 (21), 2005.74 (22), 2007.69 (23), 2015.87 (24), 2019.57 (25), and 2022.55 (26), 
where we use brackets to denote the sequence number. The accuracy of timing is typically 0.01 yr. 

Eight flares, namely the ones in 1886, 1896, 1898, 1906, 1922, 1923, 1934, and 1935, have not been properly detected due to lack of observations at those specific times. 
Furthermore, we would like to stress that there are no known flares in the historical light curve that would invalidate the above sequence. Finally, 
it should be noted that the latest thermal flare was predicted to occur during July/August 2022, at the time when OJ~287 is not observable from the Earth \citep{val07,Ram07}. 

In this paper we ask if there is any reasonable
possibility that the flare could have shifted from the unobservable to the observable part of the year from Earth's perspective. We note that \cite{val07}  
presented two slightly different precession rates for the secondary BH orbit; we refer to them as the 37.5 degree precession model and the 39.1 degree precession model.
It turned out that both these models were consistent with the available data sets of that time (the year 2006) and 
they provided similar predictions for the 
2007, 2015, and 2019 thermal flares  which are all now 
observationally verified. 

However, with the inclusion of additional data, it was realised that the 37.5 degree model does not agree with historical data. In particular, the well observed  1913 flare is problematic in the 37.5 degree model. The updated \cite{Dey18} model, after incorporating several general relativistic contributions to the BH binary dynamics, now supports the orbital precession rate of $38.62\pm0.01$ degrees per orbit.

Comparing the implications of these different precession rates, we may note that in the 39.1 degree model the flare comes at 2022.54, in the currently best model at 2022.548 \citep{Dey18}, while in the 37.5 degree precession model flare begins at 2023.13, i.e. in February 2023.

There are also astrophysical considerations that can introduce 
uncertainties in the prediction for the thermal flare arrival 
epoch, especially for the last two apastron flares of 2015 and 2022.
This is because of the possibility that the accretion disc does not stay exactly at its mean plane, but can bend slightly (of the order of 1 degree) on either side of it, due to the tidal influence of the approaching secondary SMBH. In  
the models, this was taken care by a single-valued function of the distance of the impact point from the primary SMBH \citep{val07,Dey18}. Such influences come via the parameter $t_{adv}$, which is the time difference between the epochs of the impact on the disc and on the average midplane.

Further, the delay between the impact and the start of the flare $t_{del}$ is calculated \citep{LV96} and the  difference, namely  $t_{del} - t_{adv}$, is added to the midplane crossing epoch, in order to estimate the thermal flare arrival epoch.
The need to use the parameter $t_{del}$, even though an additional parameter in the orbit solution, is a blessing in disguise, as it allows the determination of the astrophysical parameters of the disc in the standard Shakura-Sunyaev framework \citep{val19}.

From these one may calculate, e.g. the total V-band magnitude of the disc, $V \sim 19$, which means that we do not need to worry about the contribution of the disc to the total light. The faintest OJ~287 has ever been observed is at $V \sim 17.5$ \citep{tak90}.

However, there are additional difficulties with such a prescription especially when we try to compare the flares $\#\ 22$ and $26$.
Earlier numerical simulations tentatively suggested that the disc bending is quite different during the SMBH impact epochs associated with these two cases, and actually in opposite directions even though the distances of their impact sites from the centre are roughly the same \citep{val07}. The case for the 2022 disc was not properly studied so that the level of the disc at this time was essentially unknown. 

In this paper we will use previously unpublished data from these simulations,
to obtain an estimate for the disc level associated with the 
2022 BH impact.

A related problem arises from the fact that the distances of impact on the disc and on the midplane are different, when the angle of incidence is far from perpendicular,
and for the two cases of interest this angle is close to 45 degrees. Further, the impact direction is different with regard to the primary: "from inside" in 2022, and from the opposite direction, "from outside" in 2005.
These considerations suggest that we cannot simply copy the values of $t_{del}$ and $t_{adv}$ from the 2005 impact and use them in 2022 without introducing additional uncertainties.

We now move on to discuss the usual observational pattern for OJ~287. 
Typically the last optical data before the summer break are obtained in the beginning of July, while the  monitoring resumes again from the beginning of September. 
The gap in the observations, as noted earlier, arises due to 
OJ~287's small solar elongation during that window making it 
difficult to view the source from the Earth.

Thus one cannot exclude the possibility that OJ~287 had the thermal flare number 26 
during the summer break of 2022, as it was expected on July 20, 2022.
However, a slightly lower precession rate for the orbit of the secondary BH
by just $1\%$ would have shifted the flare forward in time by two weeks, which would have allowed us the opportunity to see at least the tail end of the expected thermal flare.

It should be noted  that the thermal flares do not have a counterpart in radio or X-ray wavelengths, where it is possible to get data at smaller solar elongations.

The fact that the predicted large impact flare of 2022 could not be subjected 
to multi-wavelength observational campaigns should not be too discomfiting.
This is due to the possibility that 
there may exist observational signatures associated 
with the accretion disc impact of the secondary SMBH even closer 
to the impact epoch. In what follows, we present what was known beforehand of such smaller disc impact flares \citep{val21}.

We note that the secondary SMBH impact is expected during January 2022 according to the updated \citet{Dey18} model that we refer from now onward  as the `standard model' and this is in the middle of the best observing epochs for OJ~287.

There are several observational signatures for recognising such a pre-flare,
as documented during the 2005 campaign \citep{ciprini2008}.
They include: 

(i) fast variation of polarisation similar to the main flare  \citep{val08,val19}; (ii) an exceptionally flat optical spectrum, which may be construed as a combination of the impact flare component of spectral index $\beta\sim0.75$ and the much steeper background from the jet $\beta\sim1.6$, leading to a combined colour which is much bluer than normal \citep{ciprini2008,val19}. The spectral index $\beta$ is defined in the usual manner by $F_{\nu} \sim \nu^{-\beta}$, where $F_{\nu}$ is the flux density at the frequency $\nu$.; (iii) a purely optical/UV flare with no X-ray counterpart, which implies that  the ratio $F_V/F_X$ (flux in the V-band over flux in X-rays) peaks strongly during such a flare; and (iv) there should be no radio flare associated with the optical/UV pre-flare \citep{LV96,ciprini2008}. 
We note that the pre-flares should be even better observational markers for specifying  the secondary SMBH trajectory than the big impact flares which have been used so far.

 Most of the big impact flares during the well covered portion of historical light curve, since 1970, have happened close to the pericenter. Then the big flare follows so close to the disc impact that it is not possible to see a separate pre-flare. Only four impacts have been at the apocenter part of the orbit. The impacts preceeding the big flares of 1973 and 2015 occurred in the summer time when OJ~287 was not observable from the ground. The 2005 impact was the first opportunity to study the direct emission from the impact, the 2022 flare is only the second one.

The paper is organised as follows. We begin by providing a simplified  semi-analytical formula that should allow one to obtain the first-order epochs of these bremsstrahlung flares and we refer to it as the quasi-Keplerian sequence.
Thereafter, we discuss a second-order model that is capable of producing  more accurate predictions of these impact flare arrival epochs.
A detailed description of the recent observational campaigns and how to narrow down the epoch of the recent secondary SMBH impact are presented in Section \ref{Sec3}.
The consequences of these observations which allowed us to identify 
a possible pre-flare and its implications for 
the arrival epoch of the 
traditional 2022 thermal flare are discussed in Section. \ref{Sec4}.

\section{Predicting Impact flare arrival Epochs}
\label{Sec2}
 We begin by providing a mathematical prescription for  determining a sequence of epochs that is fairly close to the one we displayed earlier. This prescription arises essentially from the celestial mechanics and general relativity considerations and is bereft of any astrophysical inputs \citep{valkar06,TG07}. We show that the 2022 flare is an essential part of the general structure, which explains the historical behaviour of flares in OJ~287.
Thereafter, we clarify why astrophysical considerations are crucial for accurately predicting the epochs of impact flare arrival times. 

\subsection{The first order ephemeris of flare times: A Quasi-Keplerian Sequence}

 We term the mathematical prescription, that provides
 a first description of the arrival epochs of impact flares as a 
 quasi-Keplerian sequence.
 This is due to the use of the classical Kepler equation, perturbed by general relativistic considerations.
 Recall that the classical Kepler equation 
  connects the eccentric anomaly $u$ to the mean anomaly $l$ \citep{valkar06}
 \begin{equation}
     l = u -e \, \sin u  \,,
 \end{equation}
 where $l = 2 \, \pi/ T_{orb}(t - t_0)$, $t$ is time, $t_0$ the perihelion time, $T_{orb}$ and $e$ are the orbital period and eccentricity, respectively, and $u$ and the phase angle $\phi$ are connected by standard formulae.
 
The quasi-Keplerian sequence, which is useful in understanding OJ~287's impact flares is characterised by an orbital period 12.13 years, eccentricity e = 0.65, forward precession $\Delta \phi = 38^{\circ}$ degrees per period and the initial angle from the pericenter to the fixed line $+1^{\circ}$ at the epoch 1910.50, one of the moments of pericenter.  Every time the particle moves over the fixed line, the phase angle of the fixed line jumps down by $\Delta \phi$, thus mimicking forward precession of the major axis of our elliptical orbit and 
 is influenced by general relativistic considerations \citep{TG07}.
The ephemeris of conjunctions is then easily calculated by using the formulae in \cite{valkar06}.
We start from the pericenter times 
\begin{equation}
T_{p}(n) = 1874.11 + 12.13 n, n=1, 2, 3...
\end{equation}
where $n$ is the orbit number.
We now invoke the Kepler Equation, written as a function of the phase angle $\phi_i(n)$ (or the true anomaly) as (\cite{valkar06}, Eqs. 3.37 \& 3.41) 
\begin{equation}
\begin{split}
T(\phi_i(n)) =  (12.13/2\pi)  (2 arctan(0.46 tan(\phi_i(n)/2)) - 0.598 tan(\phi_i(n)/2)  \\ /(1 + 0.2116 tan^2(\phi_i(n)/2))),
\end{split}
\label{Eq_KE_v}
\end{equation}

 where $\phi_i(n)$ is the phase angle at the crossing of the line of nodes \citep{valkar06}. Its values $\phi_i(n), i = -1, 0,+1$, come from the set of first flare phase angles $\phi_1(n)$, $n$ = 2,…,12, second flare phase angles $\phi_0(n)$,  $n$ = 1,…,12, and occasional third flare phase angles $\phi_{-1}(n)$, $n$ = 3,7,12:

\begin{equation}
\phi_1(n)  = (257 - 38 (n-1)) ^{\circ}, n = 2,3,8,...,12
\end{equation}
\begin{equation}
\phi _1(n) = (77 - 38 (n-1)) ^{\circ}, n = 4,...,7
\end{equation}
\begin{equation}
\phi _0(n) = (77 - 38 n) ^{\circ}, n = 1,2,8,...,12
\end{equation}
\begin{equation}
\phi_0(n)  = (257 - 38 n) ^{\circ}, n = 3,...,7
\end{equation}
\begin{equation}
\phi_{-1}(3) = 1^{\circ}
\end{equation}
\begin{equation}
\phi_{-1}(7) =  171^{\circ}
\end{equation}
\begin{equation}
\phi_{-1}(12) = 161^{\circ}
\end{equation}

where we start two orbital cycles before 1910.5 from the phase angle $(+1 + 2 \times 38)^{\circ} = 77^{\circ}$, and from the opposite phase angle $(180 + 77)^{\circ} = 257^{\circ}$ and add the third phase angle when needed. Then we get the line-crossing times

\begin{equation}
T_1(n) = T_{p}(n) - T(\phi_1(n))
\end{equation}
\begin{equation}
T_0(n) = T_{p}(n) + T(\phi_0(n))
\end{equation}
\begin{equation}
T_{-1}(n) = T_{p}(n) + T(\phi_{-1}(n))
\end{equation}

This produces a list of times with a sequence number $k = 2n-i-1$ for $k = 1,...,4, k = 2n-i$ for $k = 6,...,15$ and $k = 2n-i+1$ for $k = 16,...,26$. The sequence number $k = 5$ arises when $n = 3$ and $i = -1$. Thus the list starts $T_2(1), T_1(2), T_2(2), T_1(3), T_3(3), T_2(3), T_1(4), T_2(4), T_1(5),...$ or:
1886.49 (1), 1897.05 (2), 1898.38 (3), 1904.56 (4), 1910.51 (5), 1912.95 (6), 1922.42 (7), 1923.61 (8), 1934.24 (9), 1935.20 (10), 1945.72 (11), 1947.05 (12), 1958.12 (13), 1958.97 (14), 1964.01 (15), 1971.10 (16), 1973.04 (17), 1983.00 (18), 1984.07 (19), 1994.77 (20), 1995.77 (21), 2006.06 (22), 2007.64 (23), 2015.76 (24), 2019.57 (25), 2023.59 (26),....

A close inspection reveals that the above 
 sequence of epochs is rather close to the list of epochs that arise  from the
SMBH binary central engine description \citep{Dey18}. 
Specifically, the triplet of epochs, namely  2015.76, 2019.57 and 2023.59 in our Keplerian sequence 
closely follow the epochs  
2015.87, 2019.57 and 2022.55, that arise from 
the full mathematical solution.

 Let us emphasise that our quasi-Keplerian sequence has not been optimised in order to produce a close match between any particular flares in these two sequences. 
In principle, it should be possible to pursue it by adjusting the Keplerian parameters as such a calculation is straightforward and fully analytical.
We desist from such an exercise as we do not believe that a quasi-Keplerian model, without any elements of astrophysics, is realistic beyond producing the general flaring structure.
 
These coincidences, which are off by a year in some cases,
may be treated as an
illustrative of an underlying  perturbed Keplerian description for OJ~287.

\begin{figure*}
\includegraphics[scale=0.72]{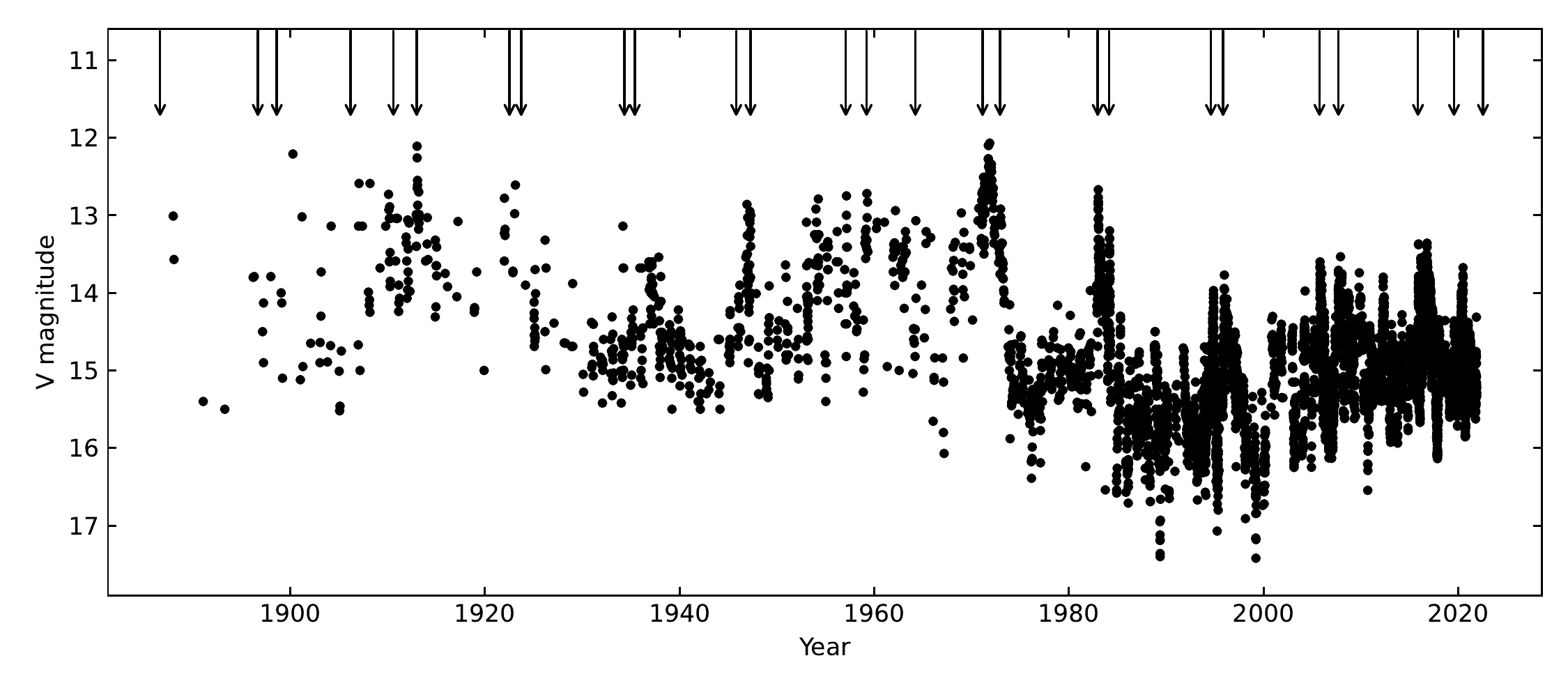}
\centering
\caption{Historical light curve of OJ~287, with the expected flare times of the standard model marked by arrows. The observations of each flare in an expanded scale, together with upper limits, are found in \citet{Dey18}.}
\label{oj287hist}
\end{figure*}

We now display 
Figure \ref{oj287hist} that shows the historical light curve where the flare epochs
are indicated by arrows.
However, we require an improved orbital description that incorporates various astrophysically relevant delays.
These considerations lead to the second order ephemeris for OJ~287.

Few comments are in order before we bring in astrophysical considerations.

Note that the requirement that phase angle $\phi_i$ should take 3 sets of values is reminiscent of the way frequencies are distributed in the GW spectrum of non-spinning BH binaries in relativistic/precessing eccentric orbits \citep{TG07}.
Recall that GWs are emitted at integer multiples of the orbital frequency for BH binaries in Newtonian eccentric orbits \citep{PM63}.
This essentially arises from the Fourier-Bessel series expansion of the Newtonian eccentric orbit in terms of the mean anomaly $l$ \citep{valkar06}.
However, each GW spectral line splits in a triplet 
when the effects of periastron advance is included \citep{TG07}.
In other words, the frequency $ f_n \rightarrow ( f_n, f_{n \pm \delta f})$ with $\delta f = 4\, \pi\,k\,f_{orb}$
and $k$ the rate of periastron advance.
This structure essentially arises from the fact that there are two timescales, that are associated with the orbital period and the periastron advance.
A similar structure in our Kepler sequence prescription naturally arises as we provide fixed angular jumps $\Delta \phi$ to the phase angle at certain fixed lines that mimic, as noted earlier, the effects of periastron advance.  We now briefly
list various astrophysical delays that should be included to generate our SMBH binary central engine description for OJ~287 and its implications.

\subsection{ Second order ephemeris for OJ~287's thermal flares and inherent astrophysical 
uncertainties }

 The above discussed quasi-Keplerian sequence allows us to pose a SMBH binary as the 
 central engine to interpret OJ~287's observations.
 Several alternate models for OJ~287 have been proposed and found to be incompatible with existing observational features of OJ~287 \citep{vil98,rie04,dey19}.
However, general relativistic effects associated with BH spins and GW emission and astrophysical considerations will have to be included into the quasi-Keplerian sequence to
obtain the standard model of \citep{Dey18}. Astrophysical considerations introduce extra parameters which are solved simultaneously with the traditional orbital elements, such as the earlier mentioned $t_{del}$ and $t_{adv}$.

 To bring in astrophysics into the above quasi-Keplerian sequence, we may identify the fixed line, present in the above sequence, with the line of nodes between the accretion disc plane and the orbital plane.
Further, we need to describe the process that generates the flares at the crossing of the line of nodes \citep{iva98}.
This leads to the time delay $t_{del}$ between the line crossing of the secondary BH and the
arrival epoch of flare \citep{LV96}.
This delay depends on where the secondary impacts the disc and can be calculated 
with the help of two parameters, the accretion rate relative to the Eddington rate, $\dot{m}$, and the viscosity parameter $\alpha$ that essentially characterise the Shakura-Sunyaev family of accretion discs
\citep{val19}.
Such considerations require  10 thermal flare arrival times, as mentioned earlier, to generate the second order ephemeris of flares that incorporate many general relativistic and 
astrophysical considerations.  For example, this leads to the epoch of 2022.55 for 
the flare $\#26$ \citep{Dey18} rather than 2023.59 that arises from the quasi-Keplerian sequence.

\begin{figure}
\includegraphics[scale=1.00]{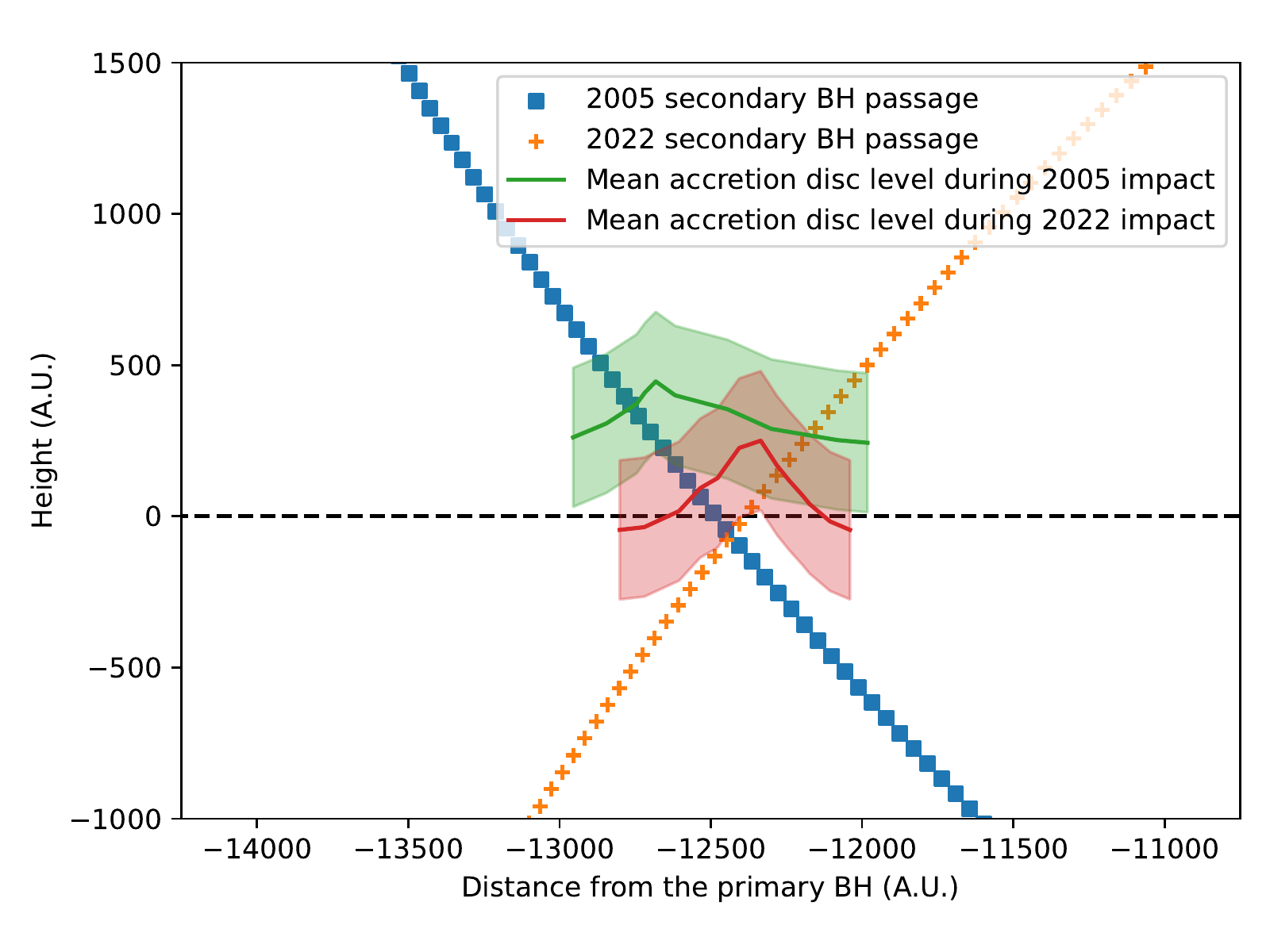}
\centering
\caption{
Accretion disc profiles around the impact sites during the 2005 and 2022 disc impact epochs. We also display the orbital path of the secondary SMBH during these times. The secondary arrives from above at both times. We see that the actual  epoch of impact depends on the disc level. Prior to the present calculation, the 2005 disc level was used for the 2022 flare arrival epoch estimates, and it 
led to July 20, 2022 as the arrival epoch of thermal flare. From here, the pre-impact disc level is estimated at $-45\pm5$ AU and the epoch of the impact flare arrival is projected within a narrow range of July 7 - July 13, 2022.}

\label{diskprofile}
\end{figure}

However, there are additional astrophysical effects that are difficult to estimate using 
semi-analytic prescriptions. This is related to the earlier mentioned $t_{adv}$ parameter that 
incorporates possible bending of the accretion disc due to the tidal 
interactions of the approaching secondary SMBH.

The 2022 impact configuration was not calculated in \cite{val07}, but since the system has a nearly perfect 109 year period, one of the 1913 impact simulations provided an excellent match of the 2022 situation, and the results of that simulation were recovered from the old archives. The simulations used non-interacting particles which were integrated along the orbit using the Aarseth-Mikkola codes \citep{aar03,mik20}. About a million disc particles were concentrated around the impact sites of 24 impacts \citep{val07}.
The 2005 impact was studied thoroughly with the full number of particles, while in a typical BH impact $\sim 10,000$ particles were used.

In Figure \ref{diskprofile}, we outline the two disc profiles, in 2005 and 2022, in the binary orbital plane. In the 2005 impact, a tidal stream forms (not illustrated in the figure), which starts from the vertex of the disc profile, and turns towards the primary BH. 
From the number of particles in that stream, we estimate a mass flux of a few solar mass per year toward the primary black hole, similar to the average accretion rate in the standard model \citep{val19}.
It is plausible that such a particle stream enters the primary BH, thereby contributing 
to its brightness, even though we have not tracked the trajectories that far.

The simulation for the 2022 disc profile reveals a similar vertex point 
though the number of particles was not high enough to observe the stream.
Using the quantitative measures to characterise such streams, as pursued in \cite{Pih13a},
we may state that the strongest stream was associated with the 2015 thermal flare while the 2005 stream was weak and the 2022 stream was even weaker.

{Extrapolating from the bright 2015 after-flares and somewhat weaker 2005 after-flares, we may 
argue that the 2022 after-flares, in the autumn of 2022, should have been rather weak. 
This is consistent with the observed low level of primary jet activity after the summer of 2022 (see Figure \ref{lc05_22} below).

}

As we have seen above, the 2022 disc level should be lower 
than the 2005 disc level with respect to the mean accretion disk as displayed in
Figure \ref{diskprofile}. 
This leads to 
$t_{adv}$  estimate that is smaller than the 2005 one by 0.04 years
and corresponding  $t_{del}$ value is also smaller by 0.07 yr. 
The latter effect arises because the point of impact of the secondary in 2022 is closer to the primary than in the 2005 impact, as is also evident from Figure \ref{diskprofile}.

Put together, the starting epoch of the 2022 big thermal flare is moved back by about
0.03 yr ( $\sim 10$ days) and in other words, the thermal flare of 2022 should have started on July 10, 2022, rather than on July 20, 2022 \citep{Dey18}.
 The uncertainties include the orbital phase uncertainty which is about $\pm3$ days, the uncertainty in the disc level in 2022, which translates to $\pm1$ day in the impact time, plus the plasma expansion time uncertainty which is more difficult to quantify. We will come back to the last item at the end of the article.

 Invoking these considerations and 
 following \cite{Dey18}, we may deduce that the secondary SMBH made its first contact with the 
 accretion disc around December 20, 2021, and entered the densest part of the disc around  January 20, 2022 (JD2459600). 
 It is plausible that the associated hot plasma from the impact would have become visible on the near side after the latter moment of time, possibly causing a short optical/UV flare \citep{val21}. 
 Therefore, an observation of such a flare can give valuable information 
 on the accretion disc position as well as on the secondary BH orbit in early 2022, thus providing a further test of the binary model of OJ~287.
 
It may be noted that \cite{val21} argued that such a flare should start around Dec 23, 2021
 by using the 2005 disc model, with a caution that the final model for the 2022 disc impact was still to be calculated.

If we impose changes in the orbital precession rate by hand, and use \cite{val07}, we find that a $1\%$ change in the precession rate moves the primary flare epoch forward by 15 days. This is fully testable in the present observing campaign of OJ~287.
In what follows, we discuss various observational efforts that were motivated by the expected pre-flare from OJ~287, as well as checking if any part of the primary flare could be seen, unlikely as it was thought to be.

\begin{figure*}
\includegraphics[angle=270,scale=0.62]{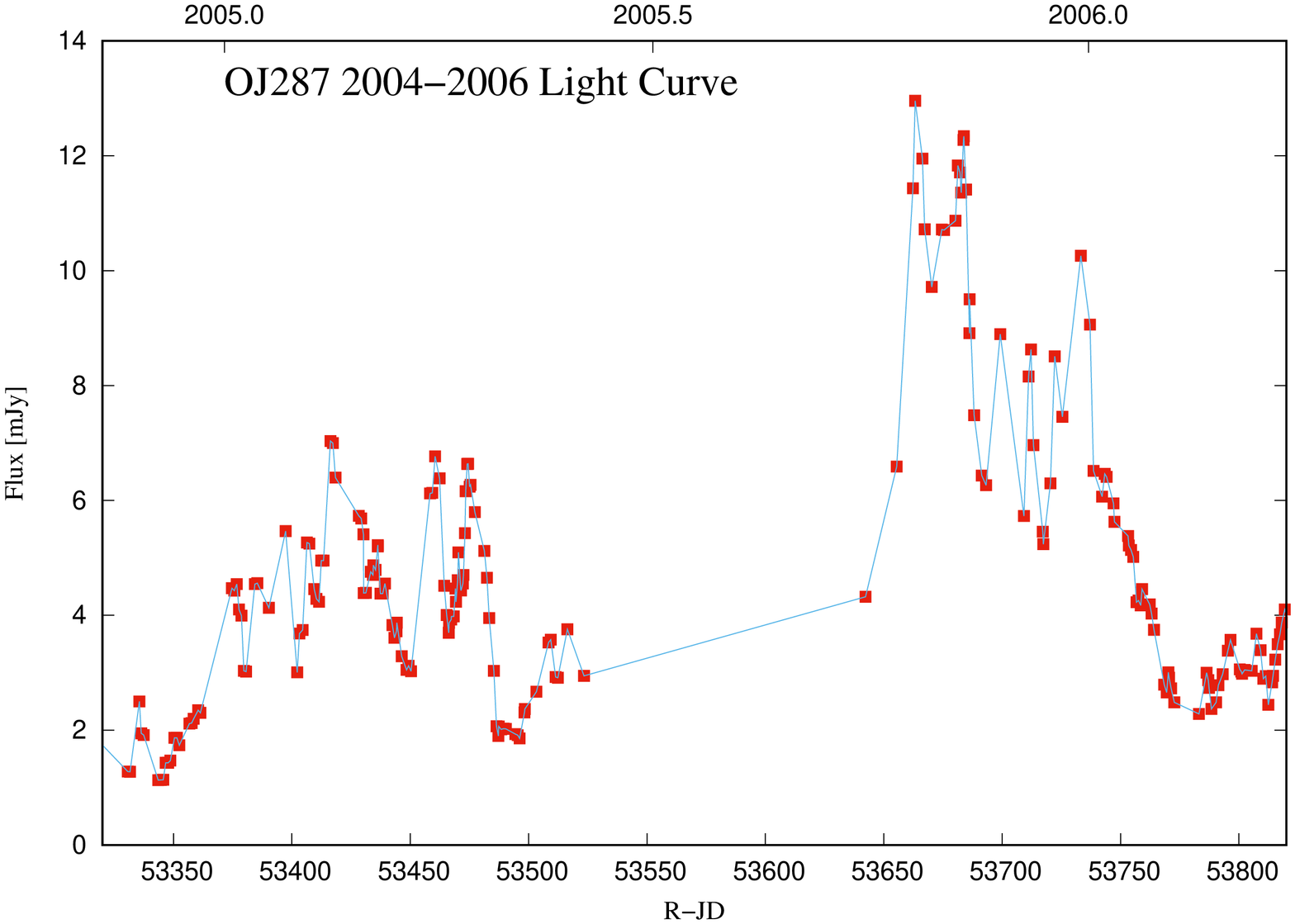}\\
\includegraphics[angle=270,scale=0.62]{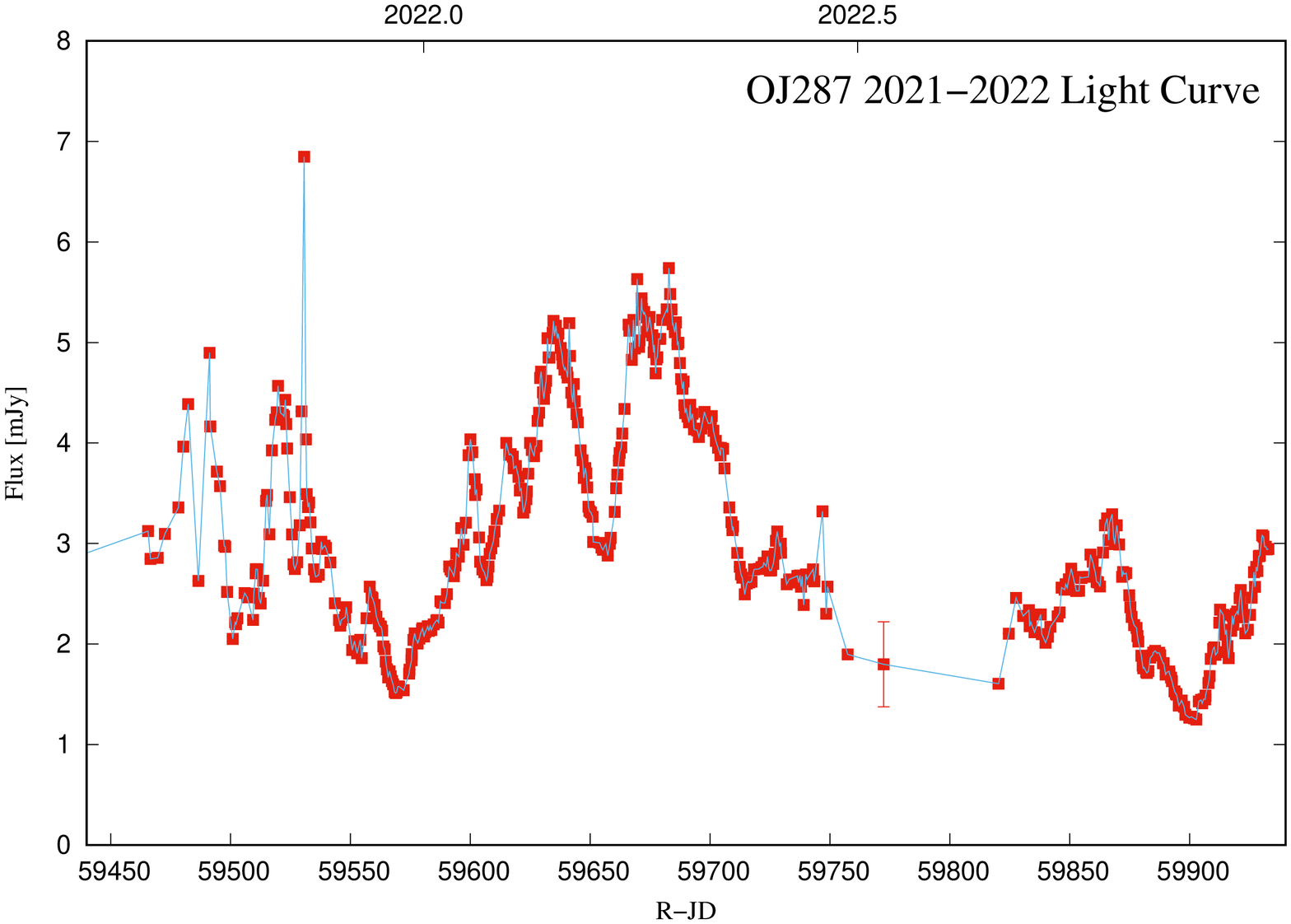}
\centering
\caption{The R-band light curves of OJ287 in the 2004/05  (top panel) and 2021/22 (bottom panel) observing seasons.The most recent observations taken after summer 2022, are also shown. In the top panel, the double-peaked thermal flare above the 6 mJy base level has the total duration which is shorter that the summer gap of the lower panel where the corresponding flare is placed by the theory. The base levels are different in the two cases, and there is no reason to expect the repeat of the 2005 base level light curve beyond the time of the thermal flare.}
\label{lc05_22}
\end{figure*}

\section{Observational Campaigns}
\label{Sec3}

OJ~287's observational campaigns during the 2021 -2022 were motivated by our 
desire to obtain any observational evidence for the predicted secondary BH impact and the 
expected thermal flare \# 26.
Astrophysical uncertainties ensured that it was not possible to obtain an expected light curve 
for flare \# 26, similar to what was done for the Eddington flare \citep{Dey18}. 
 It may be noted that astrophysical uncertainties related to the accretion disc orientation 
 suggested that the secondary BH impact may occur anytime between December, 2021 and  the beginning of March, 2022. This indicated that the occurrence of 
 the subsequent major thermal flare anytime during early June, 2022 to mid October, 2022.
 These considerations prompted us to pursue multi-wavelength photometric, spectral and polarimetric 
 observations of OJ~287 to extract any possible observational evidence for the secondary BH impact and the subsequent thermal flare during 2021-2022.

\subsection{Optical data}
\label{sec:optical}

Optical data, presented in this work, consist of older data sets, gathered in the wide band R filter \citep{val06a,wu06,ciprini2007} and a recent R filter dataset, taken within the Krakow Quasar Monitoring Program. The latter consists predominantly of observations obtained with the Skynet Telescope Robotic Network \citep{zola21}, and appended with points from other telescopes at the Observatories of Osaka, Krakow, Mt. Suhora, Ond\v{r}ejov and Jena \citep{mugrauer2010,mugrauer2016}. The location of telescopes on four continents and their redundancy allowed to achieve daily sampling, often we were able to collect data twice a day, if needed. Altogether 45217 single points have been collected since the start of the 2015/16 observing season. Binning them with half a day results in  2315 mean points. Observations discussed here cover the period from September, 2021 to December, 2022 and contain over 400 mean points, shown by red squares in the bottom panel of Figure \ref{lc05_22}.

\begin{figure}
\centering
\includegraphics[scale=0.6]{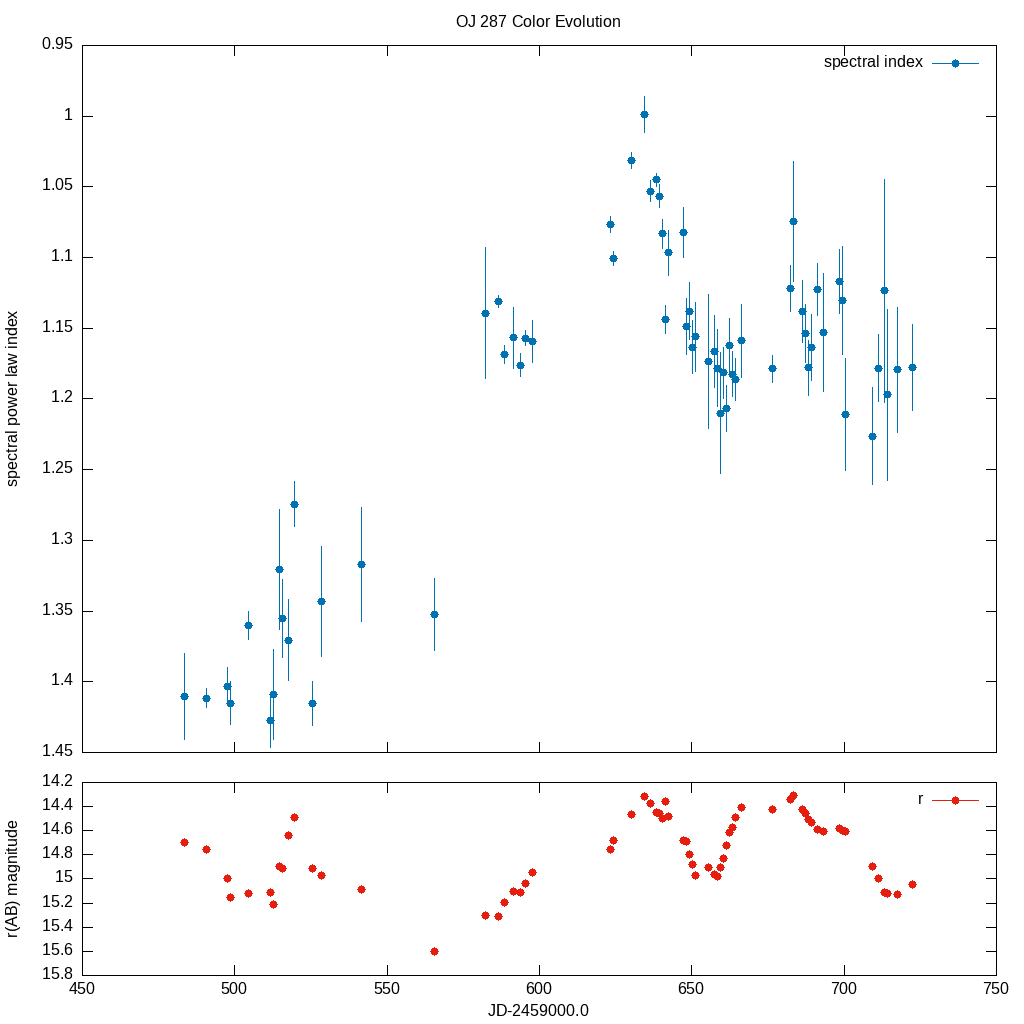}
\caption{The relation between brightness and spectral power-law colour index. The index is fitted as $F_{\nu} \sim \nu^{-\beta}$ so a higher, positive value of $\beta$ means a redder colour. The peak at $JD \sim 2459635$ corresponds to an exceptionally "blue" state.}
\label{ondrejov}
\end{figure}

Also in this campaign, photometric BVRI observations were taken with the robotic 50\,cm D50 telescope, located at the Astronomical Institute of the Academy of Sciences of the Czech Republic in Ond\v{r}ejov. The role of the telescope has been to provide complementary optical data for GRBs and other interesting high energy sources. The telescope is equipped with a low-noise emCCD camera and a set of filters (originally Johnson-Bessel BVRI, now upgraded with SDSS g',r',i',z'). The telescope observes in fully autonomous mode and the observational data are processed automatically \citep{jelinek19}. The spectral index data for the 2021/22 observing season are presented in Figure \ref{ondrejov}.

OJ~287 was well covered by optical photometry during the 2004/05 season. The points in the 2004/05 light curve are 0.01 yr averages from over 4000 single photometric observations. After a deep minimum in December 2004 there was a rather steady rise in brightness up to February 2005 maximum, followed by another maximum in April 2005.
The 2021/22 and the 2004/05 light curves are shown in Figure \ref{lc05_22}.

In addition to single channel photometry, it is important to find out how OJ~287 behaves over the whole optical spectrum. From the past experience we know that the continuum spectrum from optical to infrared has a rather constant spectral index independent of level of activity, with the BVRI spectral index around 1.35 \citep{kid18}. However, a major deviation from the relatively constant IR--optical shape happens during impact flares when the additional emission has a flat spectrum, causing the overall spectrum to flatten also \citep{val12,Laine20}. At the other end of brightness, during very deep fades the host galaxy contribution makes the IR--optical spectrum steeper \citep{val22}. For the present work it is most important to find the counterpart of the 2005 pre-flare (JD 2453474), where the BVRI spectral index decreased dramatically by more than 0.5 units in a short period of time \citep{ciprini2008}.

The bluer-when-brighter spectral trend is common in BL Lac objects, and as we said, a similar trend is also seen in OJ~287. However, for the flux range that we are discussing here, 3 - 6 mJy, the trend in OJ~287 is weak, and contributes less than 0.1 units in the spectral index  if anything at all \citep{vil10}. It will not be considered in our further discussions.

We see that the spectrum becomes unusually flat during the JD 2459638 flare. For comparison, \cite{zheng2008} find that the average spectral index of OJ~287 was 1.0 or smaller only on one occasion in the period between 1972 and 2006 which includes several intensive monitoring campaigns. This happened during the January 1973 big flare which is generally regarded as an impact flare. During 2015 - 2017 the spectral index was below or equal to 1.0 on three out of 9 occasions, and all three were high brightness states. In low brightness states the spectral index was around 1.35 \citep{Gupta17,Gupta19}. The exceptionally low value of the spectral index at the JD 2459638 flare matches well the pre-flare spectral index in 2005.

\subsection{Optical polarisation}

\begin{figure}
\includegraphics[scale=0.62,angle=270]{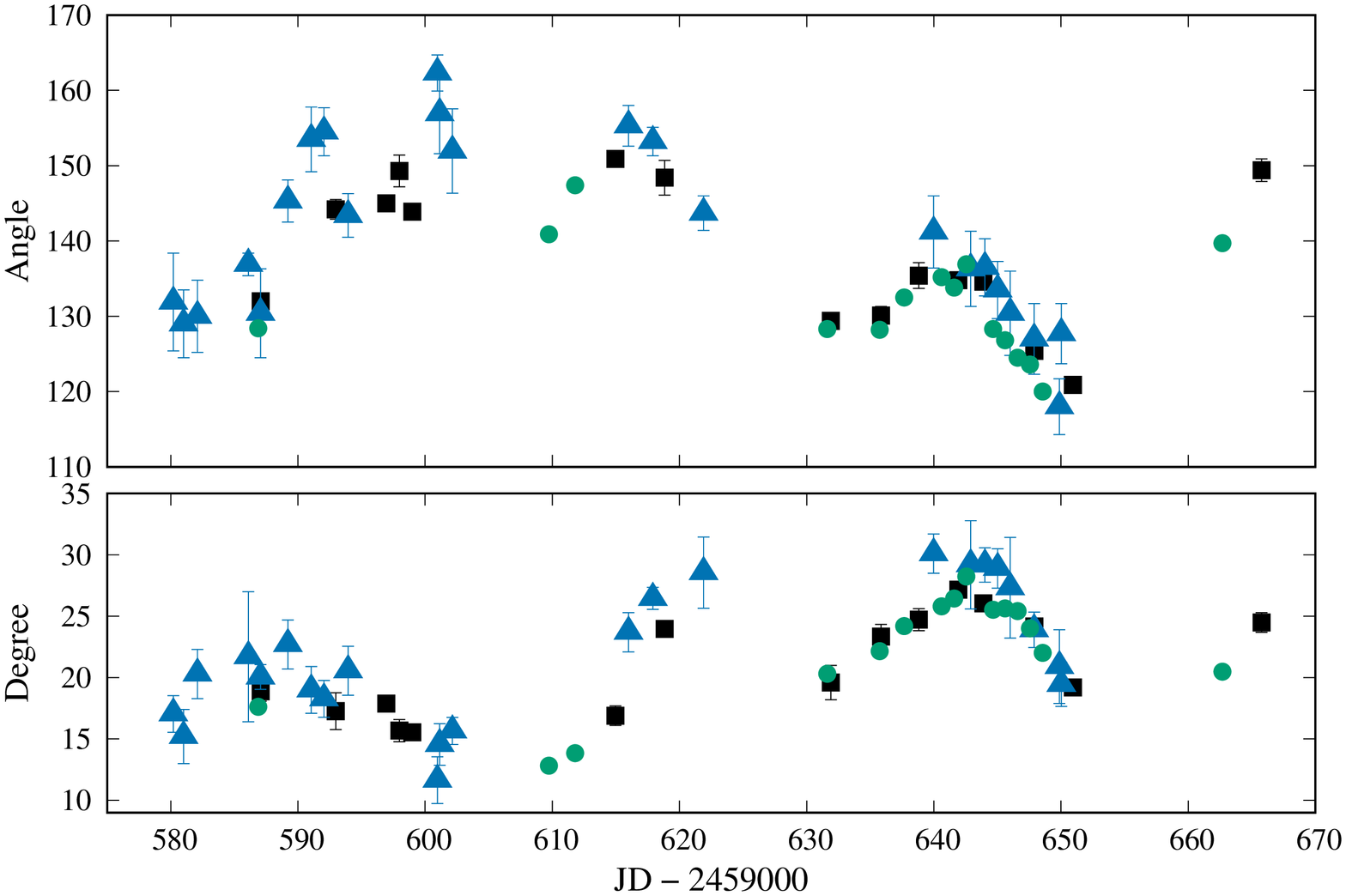}\\
\includegraphics[scale=0.62,angle=270]{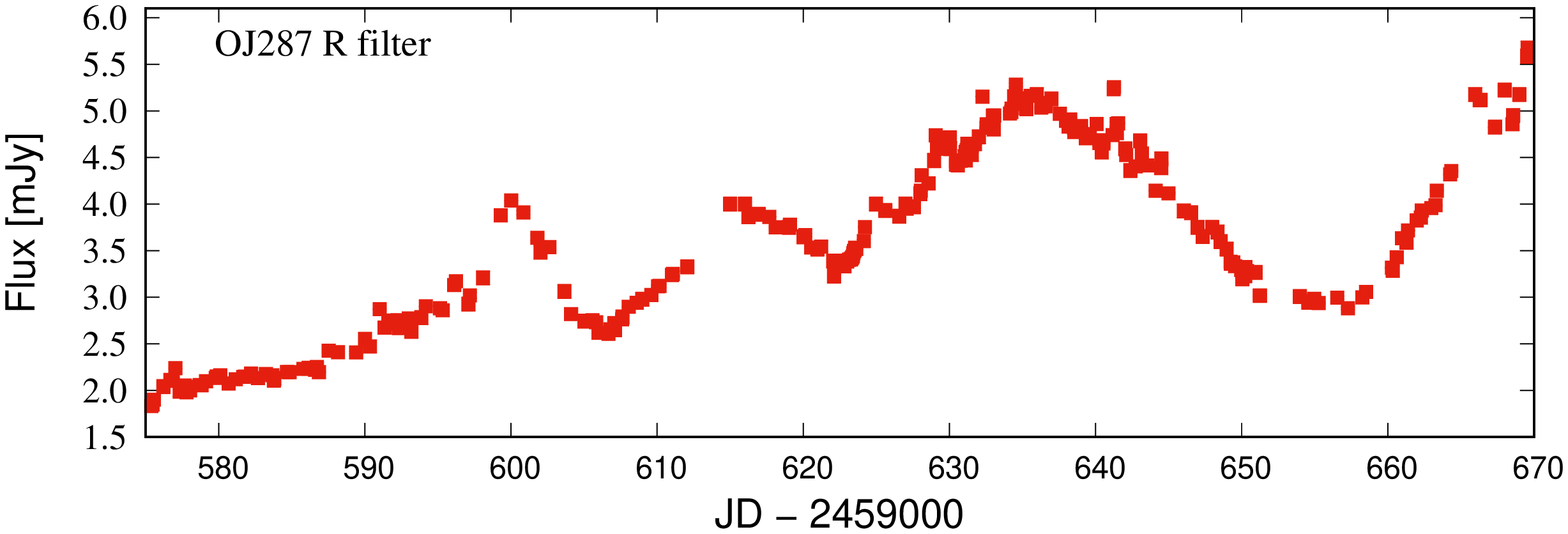}\\
\centering
\caption{The position angle of polarisation (top panel) and the degree of polarisation (middle panel) of OJ287. The points from Dipol-2 (measured by A.B.) are shown by squares, the measurements from Hiroshima (by R.I. and M.U.) are indicated by circles, and the measurements from  MOPTOP by triangles (the latter kindly provided by Helen Jermak and Callum McCall). OJ287 flux variations (R filter) in the same period are shown in the bottom panel.}
\label{polarisation}
\end{figure}

Another important aspect of the optical emission of OJ~287 is its state of polarisation which has shown large fluctuations in the past \citep{Pursimo2000,vil10}. For this campaign, polarimetric measurements were performed using the Dipol-2 polarimeter \citep{Piirola14}, mounted on the Tohoku 60cm telescope (T60) at Haleakala observatory, Hawaii. Dipol-2 is a remotely operated "double-image" CCD polarimeter, which is capable of recording polarised images in three (BVR) filters simultaneously. The innovative design of the polarimeter, where the two orthogonally polarised images of the sky overlap on the images of the source, allows us to completely eliminate the sky polarisation at the instrumental stage (even if it is variable), and to achieve unprecedentedly high, up to $10^{-5}$, accuracy of target polarimetric measurements \citep{Piirola20}. The points presented in Fig \ref{polarisation} are median values of the results from the three filters, and are nightly averages.

We also performed photopolarimetric observations of OJ~287 using TRISPEC attached to the 1.5-m “Kanata” telescope at Higashi-Hiroshima Observatory. TRISPEC is capable of simultaneous three-band (one optical and two NIR bands) imaging or spectroscopy, with or without polarimetry. TRISPEC has a CCD and two InSb arrays. Here we report R-band observations \citep{ike11}.

In addition, data from the MOPTOP - Liverpool Telescope was available to us. Each measurement consists of a frame from each of the 16 half-wave plate positions, of which frames from positions 1 - 9, 2 - 10... and 8 -16 were stacked before reduction to provide some mitigation to the loss in sensitivity from single camera operations. Observations were initially taken in just B and R filters, and later into the campaign V and I observations were added. The reduced data were then subject to a vetting procedure which performed several different quality checks on each point. The data presented in Fig \ref{polarisation} are median values at different filters and at different times within typically one day \citep{Jermak16}.

The polarisation results will be discussed in the next section. Here we may note, that the degree of polarisation rises to about 30 percent in the JD 2459638 flare, which is surprising since previously such a high degree of polarisation has been associated primarily with large flares \citep{Pursimo2000,smi09,vil10,val17} even though occasionally high degree of polarisation appears also in a low state \citep{agu11,ike11}.

\subsection{Swift data}

The X-ray band measurements of OJ~287 add an important dimension to our multi-messenger campaign.
Nasa's Neil Gehrels Swift observatory was used to study OJ~287 in the course of the project MOMO (Multiwavelength Observations and Modelling of OJ~287; \cite{Komossa2021b}). In this project, the two narrow-field telescopes aboard Swift are utilised: the UVOT and the XRT, which includes all six Swift optical and UV filters (17-600nm), and the X-rays (0.3-10keV).  The cadence ranges between typically 5 days (at inactive states) and 1 day (at outburst states or other states of particular interest). An analysis of timing and spectral properties of OJ~287 at all states of activity until January 2022 has been presented in a sequence of publications \citep{Komossa2020,Komossa2021a,Komossa2021b,Komossa2021c,Komossa2021d,Komossa2022a}. The data, mentioned here, cover the time interval 2021 October to 2022 March, and they were kindly provided to us by S. Komossa and D. Grupe. No Swift observations of OJ~287 were carried in July/August 2022.

The X-ray emission of OJ 287 is closely correlated with the optical--UV during major outbursts (most recently in 2016/17 and 2020).

\begin{figure*}
\includegraphics[width=\textwidth]{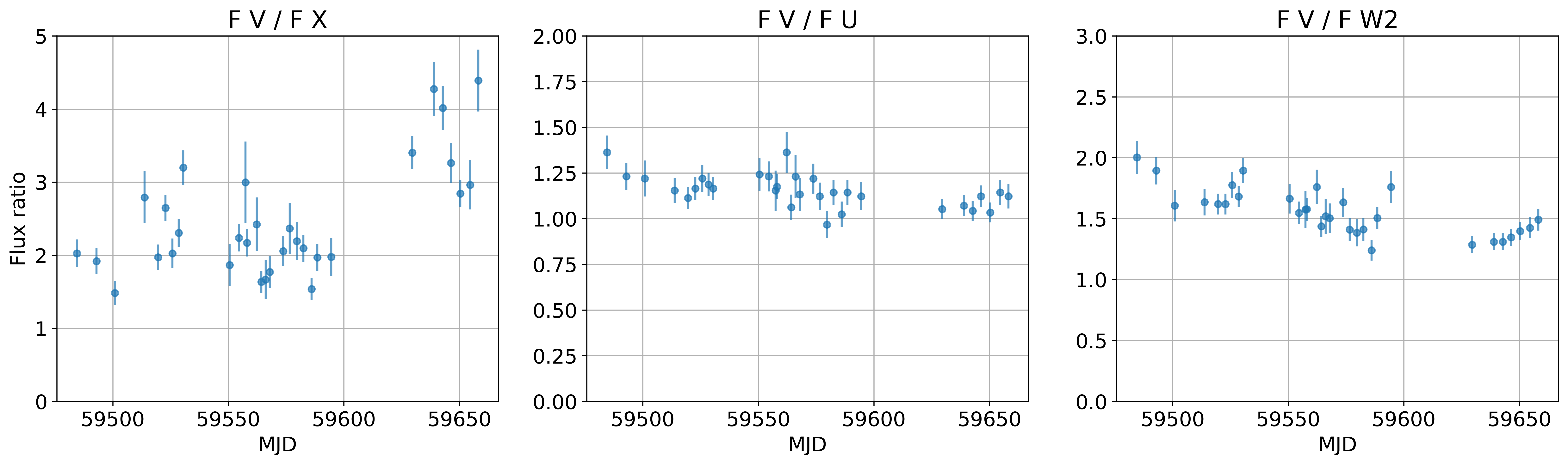}
\centering
\caption{
Spectral ratios between the visual flux $F_V$, near-ultraviolet flux $F_U$, far ultraviolet flux $F_{W2}$ and X-ray flux $F_X$, extracted from the Swift data. The flare, visible in  $F_V/F_X$ plot, occurs at $\sim$JD2459638  and this epoch is fairly close to the secondary SMBH disc impact in our description. In our model, we associate its origin to the plasma, ejected from the disc by the impact of the secondary SMBH, that may have eventually 
produced the Bremsstrahlung flare during July-August 2022. The other ratios are consistent with this being an UV dominated flare.}

\label{swift_spec}
\end{figure*}

Figure \ref{swift_spec} shows spectral ratios in selected bands in the optical, UV and X-rays during the epoch of interest between October 2021 and March 2022. We may note that the flux ratio $F_{\rm V}/F_{\rm X}$ during the peak of the JD 2459638 flare is high, actually at its highest level since mid 2020 \citep{val16,Komossa2020,Komossa2021a}, except for a short period around 2020.96, when a flare unrelated to the primary jet called "precursor" was expected \citep{Pih13a}. It is almost as high as it was during the 2005 (JD 2453474) pre-flare, which was another exceptionally high point in the $F_{\rm V}/F_{\rm X}$ light curve.

\subsection{Mets\"ahovi radio data}
Radio observations are important not because we expect anything interesting to happen at radio frequencies at the time of the BH disc impact, but exactly for the opposite reason: the radio flux should not vary greatly either at the smaller direct impact flare or during the later big flare.

We have carried out observations at the Aalto University Mets\"ahovi Radio Observatory at 37 GHz as a part of a long-standing radio monitoring programme. The average 37 GHz flux density was found to be 
$ \sim 8\pm0.5$ Jy, showing only small variability within this range during the month of the pre-flare.

Further, the radio flux follows the trend seen in the optical, with a flux decline towards the summer gap and even beyond it. 
We list here very preliminary estimates: the last measurement before the summer gap at Mets\"ahovi on July 7 
gave the flux density of 6.5 Jy, while the next measurement on August 15 gave 5.4 Jy. A detailed analysis of these radio observations will be reported elsewhere. For comparison, at the time of the 2005 pre-flare the 37 GHz flux of OJ~287 was also quite stable at $2.3\pm0.2$ Jy \citep{ciprini2008}.

\section{Conclusions and Discussion}
\label{Sec4}

 We provided an updated estimate for the epoch of the secondary BH impact on the disc of the primary in OJ~287 around  January 20, 2022. 
 This change essentially arises from our updated estimate of the accretion disc bending induced by the tidal interactions  of the approaching secondary BH \citep{val07}.
In this paper, we provided an improved estimate for the arrival epoch of 
  the 2022 Bremsstrahlung flare  from OJ~287 to be in the  July 7- July 13, 2022 window,
 which is different by about 10 days from an earlier 
 prediction, detailed in the standard orbit model of \cite{Dey18}.
 Unfortunately, both predictions are impossible to monitor from 
 the earth due to the small solar elongation of OJ~287 during the summer months.
 Interestingly, the secondary BH impact is expected to produce 
  a smaller flare of about 6 mJy several weeks after the impact, as noted in \citep{val21},
  and therefore, we explored possible observational evidence for such 
  pre-flare activities. We may note that in our  description for OJ~287 the bigger 
  thermal flare of about 12 mJy 
  is expected to occur  roughly 7.5 months after the BH impact \citep{val21}.

 The earlier observational campaigns suggest that the thermal flares typically 
 stay $\sim 35$ days above the general background variability level \citep{val11b,val16},  
 and it includes the $\sim 16$ days between the starting time and the high flaring state of these outbursts.
 Therefore, the updated description for the occurrence of the big 2022 impact flare 
 is fully consistent with the unavoidable observational gap in OJ~287's monitoring during July-August. The gap has been usually quite a bit longer than the 48 days that we had this time, as you can see by comparing the the 2005 and 2022 light curves in Figure \ref{lc05_22}.
 
The short observing gap this time was achieved because the first observations of the fall season of 2022 were carried out in Osaka already on August 28 ($R_c=15.49\pm0.09$, K.M.).  
OJ~287 was found in a low state, as shown in Figure \ref{lc05_22}. 
The last observing point of the spring season was measured as late as on July 11, 2022 in La Palma (T.P.), and OJ~287 was then also in a low state ($i = 14.8 \pm 0.25$, and with further data reduction by S.Z., giving a somewhat fainter state of R = 15.31 $\pm$ 0.26). 
These data points 
around the observational summer gap are fully consistent with the standard model of \cite{Dey18}.
 For example, if we place  the best monitored  bremsstrahlung flare  of 2015 as detailed in \cite{val16} in the above summer gap, the two observations imply the range of possible starting times of the bremsstrahlung flare from July 11 to July 26, 2022. Or you can simply take the part of the 2005 main flare from upper panel of Figure \ref{lc05_22} which is above the current base level (about 6 mJy), and place it in the summer gap of the lower panel, and it fits with even some room to manoeuvre.

 Unfortunately, no space telescope was available to monitor the predicted 2022 impact flare unlike in 2019 when the \emph{Spitzer Space Telescope} was still operational \citep{Laine20}.

 However, the main target of this campaign was the exploration of any pre-flare activities during January and February 2022, influenced by the fact 
 that the secondary impact is expected to happen during that time. 

As we pointed out in the introduction, we expected the 2022 pre-flare to be similar to the 2005 pre-flare, and in some ways also similar to the big flare that was due 6 months later. However, the pre-flare should arise from a much more compact plasma than the big flare, and therefore the physical conditions of the plasma must be different. For example, the greatly different density, temperature and magnetic flux density will necessarily cause differences in the expected radiation properties, such as the spectral index. The detailed descriptions of the plasma clouds in these two states are discussed in \cite{val19}.

One of the peculiarities of the 2022 (JD 2459638) pre-flare is the rapid variability of the degree of polarisation, especially at the time of the smaller flare of JD 2459600. In this respect the 2022 pre-flare activity has a great resemblance to what was seen in 2005 at the corresponding time \citep{vil10}, see Table 1. The range of variation in the degree of polarisation was practically the same in both cases.

Suitable models for describing the pre-flares may be found from \cite{van71} who describes the radiation of a uniformly expanding bubble. For example, the flare in 3C~273 in 1967 shows a fairly simple brightness profile, while there is plenty of structure in both the position angle and the degree of polarisation of the flare. Early in the flare the degree of polarisation goes down sharply but then quickly recovers. At the same time there are large swings in the position angle of polarisation. In our case, we have to add the base level component which may have its own more slowly changing polarisation properties.

It may be noted that the position angle of the primary jet, as determined from certain jet models \citep{Dey21} as well as from VLBI observations \citep{Gom22}, roughly agrees with the optical polarisation position angle reported here. These models predict 
 $PA = 123 - 128^{\circ}$ \citep{val21} while the recent observations at quiescent times  provide
 $PA \sim 125\pm15^{\circ}$.

In radio wavelengths, no coincidental flares corresponding to the two optical pre-flares
have been detected. The variability percentage in both cases was rather similar, see Table 1. During the summer gap, which was partially covered by our radio observations, there was no indication of radio flares, and none was expected.

In X-rays (Figure.~\ref{swift_spec})
 we infer the occurrence of a  
 flare with a very prominent $F_V/F_X$ peak around 
 JD2459638. Its $F_V/F_X$ ratio is somewhat smaller than what was observed in the 2005 pre-flare \citep{ciprini2008,Komossa2021d}, see Table 1. It is understandable, since in 2005 the optical flare rose higher above the background level than in 2022. However, this statement is somewhat conditional on getting the correct divide between the background and the flare components. Table 1 lists the flare contribution, if the background was 2 mJy at both instances.

Also, the behaviour of the optical spectral index is exceptional during the 2005 and 2022 pre-flares. The spectral index at the early 2022 flux level should be around 1.35 \citep{zheng2008}, but actually it is as low as 1.0. 
We gather from past observations that 
the only other epoch when OJ~287  has had such a low spectral index during a low activity state,  was at the 2005 (JD 2453474) pre-flare. 
Especially when comparing the BVRI spectral index data, it becomes evident  that the 2022 (JD 2459638) pre-flare
is the counterpart of the 2005 (JD 2453474) pre-flare with regard to the spectrum, see Table 1.

The spectral index as low as 1.04 measured in the 2005 pre-flare \citep{ciprini2008}, is consistent with adding a flat component of spectral index $\beta\sim0.75$ on top of the background jet emission of much higher spectral index. 
A similar statement about the 2022 (JD 2459638) pre-flare is possible
where the background radiation values are somewhat different during 2022 than in 2005. 
These observed spectral indices agree with the typical spectral index in a transparent synchrotron source \citep{pac70}.

The models of \cite{van71} also tell us what should happen to the spectral index during the flare. Early on the bubble is optically thick and it has an inverted spectrum, $\beta \leq$ 0. The flare has not started yet at this stage. When the source becomes optically thin, the brightness goes up sharply, and the spectral index becomes $\beta \geq$ 0.5. Therefore, the combined spectral index of the base level ($\beta \sim 1.5$) and the flare ($\beta \sim 0.5$) should be $\beta \sim 1.0$, when the base and the flare make about equal flux contributions.

\cite{Komossa2023} expect the combined spectral index to be $\beta \sim 0.2$ which implies that the flare component should have a spectral index $\beta \sim - 1.0$, i.e. it would have an inverted spectrum. This is not possible in the usual expanding plasma cloud models. 

The pre-flare properties in 2005 and 2022 are summarised in Table 1. There we also note (last column) that the two flares arise at the same orbital phase $\Phi$ with respect to the accretion disk. The timing is such that the emerging plasma cloud, released from the disc by the BH impact, has just come to the surface of the disc when the flare comes to its peak. Even though the observed plasma cloud comes toward us, the BH recedes to the other side of the disc, their outward speeds are about the same \citep{LV96,iva98} and therefore the BH distance from the disc measures also the corresponding distance for the plasma cloud.

\begin{table}
\begin{center}
\caption{Comparison of 2005 and 2022 pre-flares. The columns are (1): year, (2) Julian Day of the peak, (3) size of flare in mJy, (4) $F_V/F_X$, (5) Variation of percentage polarisation, (6) Variation of radio flux, (7) spectral index, (8) orbital phase angle at the peak of the flare.}
\label{tab:dataAMAT}
\begin{tabular}{lccccccc}
\hline
$year$&$JD$&  $R (mJy)$&$F_V/F_X$&$\Delta P\%$&$\Delta F_R (\%)$&$\beta$&$ \Phi (deg)$    \\
\hline
  2005&  2453474& 4.4 &6.3  $\pm$ 0.7& 17 & 9 &1.04 $\pm$ 0.02&2.0 $\pm$ 0.2\\
  2022&  2459638& 3.0 &4.3  $\pm$ 0.4& 18 & 6 &1.00 $\pm$ 0.02&2.0 $\pm$ 0.2\\
\hline
\end{tabular}
\end{center}
\end{table}

The identification of the 2005 pre-flare also helps us to check the model for the 2005 main flare. From the pre-flare we can estimate the time of the disc impact in 2005, and since we have observed the 2005 main flare, we are able to measure the time difference between them, $t_{del}$. The result agrees with \cite{Dey18} within 3 days. Note that the latter model depends only on the difference $t_{del} - t_{adv}$, not on the values of $t_{del}$ and $t_{adv}$ individually, so that this is the first time that $t_{del}$ has been determined independently.
We would like to emphasize that it is the additional spectral index data 
that gave us the confidence to relate the pre-flares in the 2005 and 2022 light curves.
If we were to associate 
the observed flare at $\sim$ JD 2459675 with the 2005 pre-flare, the arrival of the 2022 impact flare should also be shifted to around October 10, 2022.

When this observing campaign started, we did not know the geometrical configuration of the 2022 disc impact. So why does it matter? The orbit model of OJ~287 relies partly on knowing the impact geometry during those impacts which are used for the mathematical orbit solution. Therefore it is important to get an independent verification of the disc levels. This has now been done in two ways, by going back to simulation archives, and by comparing 2005 and 2022 multi-messenger light curves of OJ~287. Both methods give the same result within their associated uncertainties. This gives us confidence that, for example, the spin of the primary BH, which is strongly influenced by the timing of the 2015 flare, is correctly determined within the accuracy of the published error limits \citep{val16}.

It is also important to verify that the accretion disc model used for the OJ~287 work is correct and self-consistent. The model comes from the Shakura-Sunyaev family of accretion disc models \citep{sha73,sak81,ste84,LV96}, and has parameters $\dot{m} \sim 0.08$ and $\alpha \sim 0.26$ \citep{val19}. The value of $t_{del}$ is most sensitive to these parameters. We have confirmed our previous value $t_{del}$ at the 2005 disc impact, which gives us confidence on using these parameters in our disc model. These values place the model in the standard sequence of thin disc models \citep{che95,zdz98}, and clearly outside the range of models like ADAF. The lack of a thin disc in the latter models would make them unsuitable for modelling OJ~287 \citep{liu22}.

\section*{Acknowledgements}
Data from the Steward Observatory spectropolarimetric monitoring project were used. This program is supported by Fermi Guest Investigator grants:       NNX08AW56G, NNX09AU10G, NNX12AO93G, and NNX15AU81G. We are grateful to S. Komossa and D. Grupe for providing information on the Swift data, presented in this paper, that comes from their MOMO observing programme, and for valuable discussions. We also thank Helen Jermak and Callum McCall for providing polarisation data prior to publication. This work was partly funded by NCN grant No. 2018/29/B/ST9/01793 (SZ) and JSPS KAKENHI grant No. 19K03930 (KM). Part of this work is based on archival data, software or online services, provided by the Space Science Data Center, SSDC, of the Italian Space Agency (Agenzia Spaziale Italiana, ASI). SC acknowledges support by ASI through contract ASI-INFN 2021-43-HH.0 for SSDC, and Instituto Nazionale di Fisica Nucleare (INFN). This paper makes use of data obtained at Mets\"ahovi Radio Observatory, operated by Aalto University in Finland. RH acknowledges the EU project H2020 AHEAD2020, grant agreement 871158, and internal CTU grant SGS21/120/OHK3/2T/13. ACG is partially supported by Chinese Academy of Sciences (CAS) President's International Fellowship Initiative (PIFI) (grant no. 2016VMB073). MJV acknowledges a grant from the Finnish Society for Sciences and Letters.

\section*{Data Availability}
The data published in this paper are available on reasonable request from the authors.

\bsp    
\label{lastpage}

\begin{thebibliography}{999}
\bibitem[Aarseth(2003)]{aar03}{Aarseth, S.J.} \textit{Gravitational N-Body Simulations, Cambridge, UK: Cambridge University Press} \textbf{2003}
\bibitem[Agudo et al.(2011)]{agu11} {Agudo, I., Jorstad, S.G., Marscher, A.P., Larionov, V.M., Gomez, J.L., L\"ahteenm\"aki, A., Gurwell, M., Smith, P.S., Wiesemeyer, H., Thum, C. et al.}, \textit{ApJL} \textbf{2011}, \textit{726}, L13
\bibitem[Begelman et al.(1980)]{BBR80} {Begelman, M.C., Blandford, R.D. \& Rees, M.J.}, \textit{Nature} \textbf{1980}, \textit{287}, 307
\bibitem[Bon et al.(2016)]{Bon2016} {Bon, E., et al.}, \textit{ApJS} \textbf{2016}, \textit{225}, 29
\bibitem[Burke-Spolaor et al.(2018)]{bur18} {Burke-Spolaor, S., Blecha, L., Bogdanovic, T., Comerford, J.M., Lazio, J., Liu, X., Maccarone, T.J., Pesce, D., Shen, Y.\& Taylor, G.}, \textit{ASPC} \textbf{2018}, \textit{517}, 677
\bibitem[Charisi et al.(2016)]{charisi2016} {Charisi, M., Bartos, I., Haiman, Z., Price-Whelan, A.M.,Graham, M.J., Bellm, E.C., Laher, R.R. \& Marka, S.}, \textit{MNRAS} \textbf{2016}, \textit{463}, 2145
\bibitem[Chen et al.(1995)]{che95} {Chen, X., Abramowicz, M.A., Lasota, J.-P., Narayan, R. \& Yi, I.}, \textit{ApJ} \textbf{1995}, \textit{443}, L61
\bibitem[Ciprini et al.(2007)]{ciprini2007} { Ciprini, S., Raiteri, C. M., Rizzi, N., Agudo, I., Foschini, L., Fiorucci, M., Takalo, L. O., Villata, M., Ostorero, L., Sillanp\"a\"a et al.}, \textit{MmSAI} \textbf{2007}, \textit{78}, 741
\bibitem[Ciprini \& Rizzi(2008)]{ciprini2008} { Ciprini, S.\& Rizzi, N.}, \textit{in Proc.
of Science, PoS(BLAZARS2008)} \textbf{2008}, \textit{63}, 30 
\bibitem[Dey et al.(2018)]{Dey18} {Dey, L., Valtonen, M.J., Gopakumar, A., Zola, S., Hudec, R., Pihajoki, P., Ciprini, S., Matsumoto, K., Sadakane, K., Kidger, M. et al. }, \textit{ApJ} \textbf{2018}, \textit{866}, 11
\bibitem[Dey et al.(2019)]{dey19} {Dey, L., Gopakumar, A., Valtonen, M., Zola, S., Susobhanan, A., Hudec, R., Pihajoki, P., Pursimo, T., Berdyugin, A., Piirola, V. et al. }, \textit{Universe} \textbf{2019}, \textit{5}, 108
\bibitem[Dey et al.(2021)]{Dey21} {Dey, L., Valtonen, M.J., Gopakumar, A., Lico, R., Gomez, J., Susobhanan, A., Komossa, S. \& Pihajoki, P. }, \textit{MNRAS} \textbf{2021}, \textit{503}, 4400-4412
\bibitem[Gómez et al.(2022)]{Gom22} { Gómez, J. L., Traianou, E., Krichbaum, T. P., Lobanov, A. P., Fuentes, A., Lico, R., Zhao, G.-Y., Bruni, G., Kovalev, Y. Y., Lähteenmäki, A. et al. }, \textit{ApJ} \textbf{2022}, \textit{924}, 122
\bibitem[Graham et al.(2015)]{Graham2015} {Graham, M.~J., Djorgovski, S.~G., Stern, D., Glikman, E., Drake, A.J., Mahabal, A.A., Donalek, C., Larson, S. \& Christensen, E.} ,\textit{Nature} \textbf{2015}, \textit{518}, 74
\bibitem[Gupta et al.(2017)]{Gupta17}{Gupta, A.~C., Agarwal, A., Mishra, A., Gaur, H., Wiita, P.~J., Gu, M.~F., Kurtanidze, O.~M.,  Damljanovic, G., Uemura, M., Semkov, E., et al.}, \textit{MNRAS} \textbf{2017}, \textit{465}, 4423
\bibitem[Gupta et al.(2019)]{Gupta19}{Gupta, A.C., Gaur, H., Wiita, P.J., Pandey, A., Kushwaha, P., Hu, S.M., Kurtanidze, O.M., Semkov, E., Damljanovic, G., Goyal, A. et al}, \textit{AJ} \textbf{2019}, \textit{157}, 95
\bibitem[Hudec et al.(2013)]{Hudec2013} {Hudec R., Basta, M., Pihajoki, P. \& Valtonen, M.}, \textit{A\&A} \textbf{2013}, \textit{559}, 20
\bibitem[Iguchi et al.(2010)]{Iguchi2010} {Iguchi, S., Okuda, T., \& Sudou, H.}, \textit{ApJ} \textbf{2010}, \textit{724}, L166
\bibitem[Ikejiri et al.(2011)]{ike11}{Ikejiri, Y., Uemura, M., Sasada, M., Ito, R., Yamanaka, M., Sakimoto, K., Arai, A., Fukazawa, Y., Ohsugi, T., Kawabata, K. S., et al.}, \textit{PASJ} \textbf{2011}, \textit{63}, 639
\bibitem[Ivanov et al.(1998)]{iva98} {Ivanov, P.B., Igumenshchev, I.V. \& Novikov, I.D.}, \textit{ApJ} \textbf{1998}, \textit{507}, 131-144
\bibitem[Jel\'{\i}nek et al.(2019)]{jelinek19}{{Jel{\'\i}nek}, Martin and {Kann}, David A. and {{\v{S}}trobl}, Jan and {Hudec}, Ren{\'e}}, \textit{AN}, \textbf{2019} , \textit{340}, 622-628
\bibitem[Jermak et al.(2016)]{Jermak16} {Jermak H., Steele, I. A., Smith R. J.}, \textit{SPIE} \textbf{2016}, \textit{9908}, 99084I
\bibitem[Kaur et al.(2017)]{kau17} {Kaur, N., Sameer, Baliyan, K.S. \& Ganesh, S.}, \textit{MNRAS} \textbf{2017}, \textit{469}, 2305
\bibitem[Kidger (2007)]{kid07} {Kidger, M.} Cosmological Enigmas: Pulsars, Quasars, and Other Deep-Space Questions. \textit{The Johns Hopkins University Press} \textbf{2007}
\bibitem[Kidger et al.(2018)]{kid18} {Kidger, M., Zola, S., Valtonen, M., L\"ahteenm\"aki, A., J\"arvel\"a, E., Tornikoski, M., Tammi, J., Liakos, A. \& Poyner, G.}, \textit{A\&A} \textbf{2018}, \textit{610}, A74
\bibitem[Komossa \& Zensus(2016)]{kz2016} {Komossa S., Zensus J.A.}, \textit{IAUS} \textbf{2016}, \textit{312}, 13 
\bibitem[Komossa et al.(2020)]{Komossa2020} {Komossa S., Grupe D., Parker M.L., Valtonen M. J., Gómez J.L., Gopakumar A. \& Dey L.}, \textit{MNRAS} \textbf{2020}, \textit{498}, L35 
\bibitem[Komossa et al.(2021a)]{Komossa2021a} {Komossa S., Grupe D., Parker M.L., Gómez J.L., Valtonen M.J., Nowak M.A., Jorstad S.G., Haggard D., Chandra, S., Ciprini, S. et al.}, \textit{MNRAS}, \textbf{2021a}, \textit{504}, 5575
\bibitem[Komossa et al.(2021b)]{Komossa2021b} {Komossa S., Grupe D., Kraus A., Gallo L.C., Gonzalez A.G., Parker M.L., Valtonen M.J., Hollett A.R., Bach, U., Gomez, J.L. et al.}, \textit{Universe} \textbf{2021b}, \textit{7}, 261
\bibitem[Komossa et al.(2021c)]{Komossa2021c} {Komossa S., Ciprini S., Dey L., Gallo L.C., Gomez J.L., Gonzalez A., Grupe D., Kraus A., Laine, S., Parker, M.L. et al.}, \textit{Publ. Astron. Obs. Belgrade} \textbf{2021c}, \textit{100}, 29-42
\bibitem[Komossa et al.(2021d)]{Komossa2021d} {Komossa S., Grupe D., Gallo L.C., Gonzalez A., Yao S., Hollett A.R., Parker M.L. \& Ciprini S.}, \textit{ApJ} \textbf{2021d}, \textit{923}, 51 
\bibitem[Komossa et al.(2022a)]{Komossa2022a} {Komossa S., Grupe D., Kraus, A., Gonzalez, A., Gallo, L.C., Valtonen, M.J., Laine, S., Krichbaum, T.P., Gurwell, M.A., Gómez, J.L. et al.}, \textit{MNRAS} \textbf{2022a}, \textit{513}, 3165 
\bibitem[Komossa et al.(2023)]{Komossa2023} {Komossa S., Groupe, D., Kraus, A., Gurwell, M. A., Haiman, Z., Liu, F. K., Tchekhovskoy, A., Gallo, L. C., Berton, M., Blandford, R., et al.}, \textit{MNRAS} \textbf{2023}, in press (arXiv:2302.11646) 
\bibitem[Koss et al.(2023)]{Koss23}{Koss, M.J., Treister, E., Kakkad, D., Casey-Clyde, J.A., Kawamuro, T., Williams, J., Foord, A., Trakhtenbrot, B., Bauer, F.E., Privon, G.C. et al.}, \textit{ApJL}, \textbf{2023}, \textit{942}, L24
\bibitem[Laine et al.(2020)]{Laine20} {Laine, S., Dey, L., Valtonen, M., Gopakumar, A., Zola, S., Komossa, S., Kidger, M., Pihajoki, P., Gomez, J.L., Caton, D. et al.}, \textit{ApJL} \textbf{2020}, \textit{894}, L1
\bibitem[Lainela et al.(1999)]{lai99} {Lainela, M., Takalo, L.O., Sillanp\"a\"a, A., Pursimo, T., Nilsson, K., Katajainen, S., Tosti, G., Fiorucci, M. et al.}, \textit{ApJ} \textbf{1999}, \textit{521}, 561
\bibitem[Lehto \& Valtonen(1996)]{LV96} {Lehto, H.J., \& Valtonen, M.J.}, \textit{ApJ} \textbf{1996}, \textit{460}, 207
\bibitem[Liu et al.(2014)]{Liu2014} {Liu, F.~K., Li, S., \& Komossa, S.}, \textit{ApJ} \textbf{2014}, \textit{786}, 103
\bibitem[Liu \& Qiao(2022)]{liu22}{Liu, B.~F. \& Qiao, E.}, \textit{Science} \textbf{2022}, \textit{25}, 103544
\bibitem[Mikkola(2020)]{mik20} {Mikkola, S.}, \textit{Gravitational Few-Body Dynamics: A Numerical Approach. Cambridge University Press, UK} \textbf{2020}
\bibitem[Mikkola \& Valtonen(1992)]{mik92} {Mikkola, S. \& Valtonen, M.J.}, \textit{MNRAS} \textbf{1992}, \textit{259}, 115
\bibitem[Milosavljevic \& Merritt(2001)]{mil01} {Milosavljevic, M. \& Merritt, D.}, \textit{ApJ} \textbf{2001}, \textit{563}, 34
\bibitem[Mugrauer(2016)]{mugrauer2016} {Mugrauer, M.}, \textit{AN} \textbf{2016}, \textit{337}, 226
\bibitem[Mugrauer \& Berthold(2010)]{mugrauer2010} {Mugrauer, M. \& Berthold, T.}, \textit{AN} \textbf{2010}, \textit{331}, 449
\bibitem[Nilsson et al.(2010)]{nil10} {Nilsson, K., Takalo, L.O., Lehto \& Sillanp\"a\"a, A.}, \textit{A\&A} \textbf{2010}, \textit{516}, A60
\bibitem[O'Neill et al.(2022)]{ONeill22}{O'Neill, S., Kiehlmann, S., Readhead, A.C.S., Aller, M.F., Blandford, R.D., Liodakis, I., Lister, M.L., Mr\'oz, P., O'Dea, C.P., Pearson, T.J., et al.}, \textit{ApJL} \textbf{2022}, \textit{926}, L35
\bibitem[Pacholczyk (1970)]{pac70} {Pacholczyk, A.G.}, \textit{Radio Astrophysics. W.H.Freeman and Company, San Francisco} \textbf{1970}, 140
\bibitem[Peters \& Mathews(1963)]{PM63} {Peters, P.C. \& Mathews, J.}, \textit{PhysRev} \textbf{1963}, \textit{131}, 435
\bibitem[Pihajoki et al.(2013)]{Pih13a} {Pihajoki, P., Valtonen, M., Zola, S., Liakos, A., Drozdz, M., Winiarski, M., Ogloza, W., Koziel-Wierzbowska, D., Provencal, J., Nilsson, K. et al.}, \textit{ApJ} \textbf{2013}, \textit{764}, 5
\bibitem[Piirola et al.(2014)]{Piirola14} {Piirola, V., Berdyugin, A. \& Berdyugina, S.}, \textit{SPIE} \textbf{2014}, \textit{9147}, 91478I
\bibitem[Piirola et al.(2020)]{Piirola20} {Piirola, V., Berdyugin, A., Frisch, P. C., et al.}, \textit{A\&A} \textbf{2020}, \textit{635}, 46
\bibitem[Pursimo et al.(2000)]{Pursimo2000} {Pursimo T., Takalo, L.O., Sillanp\"a\"a, A., Kidger, M., Lehto, H.J., Heidt, J., Charles, P.A., Aller, H., Aller, M., Beckmann, V. et al.}, \textit{A\&AS} \textbf{2000}, \textit{146}, 141-155
\bibitem[Quinlan(1996)]{qui96} {Quinlan, G.D.}, \textit{NewAstr} \textbf{1996}, \textit{1}, 35
\bibitem[Rampadarath et al.(2007)]{Ram07} {Rampadarath, H., Valtonen, M.J. \& Saunders, R.}, \textit{The Central Engine of Active Galactic Nuclei, ASP Conference Series, Eds. L. C. Ho \& J.-M. Wang} \textbf{2007}, \textit{373}, 243-244
\bibitem[Rieger(2004)]{rie04}  {Rieger, F.M.}, \textit{ApJ} \textbf{2004}, \textit{615}, L5-L8
\bibitem[Sakimoto \& Coroniti(1981)]{sak81}  {Sakimoto, P.J. \& Coroniti, F.V.}, \textit{ApJ} \textbf{1981}, \textit{247}, 19
\bibitem[Shakura \& Sunyaev(1973)]{sha73}  {Shakura, N.I. \& Sunyaev, R.A.}, \textit{A\&A} \textbf{1973}, \textit{24}, 337
\bibitem[Sillanp\"a\"a et al.(1985)]{sil85} {Sillanp\"a\"a, A., Teerikorpi, P., Haarala, S., Korhonen, T., Efimov, I. S., \& Shakhovskoi, N. M.}, \textit{A\&A} \textbf{1985}, \textit{147}, 67-70
\bibitem[Sillanp\"a\"a et al.(1988)]{sil88} {Sillanp\"a\"a, A., Haarala, S., Valtonen, M.J., Sundelius, B. \& Byrd, G.G.}, \textit{ApJ} \textbf{1988}, \textit{325}, 628
\bibitem[Sillanp\"a\"a et al.(1996a)]{sil96a} {Sillanp\"a\"a, A., Takalo, L.O., Pursimo, T., Lehto, H.J., Nilsson, K., Teerikorpi, P., Hein\"am\"aki, P., Kidger, M., de Diego, J.A., Gonzalez-Perez, J.N. et al.}, \textit{A\&A} \textbf{1996}, \textit{305}, L17
\bibitem[Sillanp\"a\"a et al.(1996b)]{sil96b} {Sillanp\"a\"a, A., Takalo, L.O., Pursimo, T., Nilsson, K., Hein\"am\"aki, P., Katajainen, S., Pietil\"a, H., Hanski, M., Rekola, R., Kidger, M. et al.}, \textit{A\&A} \textbf{1996}, \textit{315}, L13-L16
\bibitem[Sitko \& Junkkarinen(1985)]{sit85} {Sitko, M.L., \& Junkkarinen, V.~T.}, \textit{PASP} \textbf{1985}, \textit{97}, 1158-1162
\bibitem[Smith et al.(1985)]{smi85} {Smith, P.S., Balonek, T.J., Heckert, P.A., Elston, R. \& Schmidt, G.~D.}, \textit{AJ} \textbf{1985}, \textit{90}, 1184-1187
 \bibitem[Smith et al.(2009)]{smi09} {Smith, P.S., Montiel, E., Rightley, S., Turner, J., Schmidt, G.D., \& Jannuzi, B.T.}, \textit{Fermi Symposium, eConf Proceedings C091122} \textbf{2009}, \textit{arXiv:0912.3621, 2009} 
\bibitem[Stella \& Rosner(1984)]{ste84} {Stella, L. \& Rosner, R.}, \textit{ApJ} \textbf{1984}, \textit{277}, 312 
\bibitem[Takalo et al.(1990)]{tak90} {Takalo, L. O., Kidger, M., de Diego, J. A., Sillanp\"a\"a, A., Piirola, V., Ter\"asranta, H.}, \textit{A\&AS} \textbf{1990}, \textit{83}, 459 
\bibitem[Tessmer \& Gopakumar (2007)]{TG07} {Tessmer, M. \& Gopakumar, A.}, \textit{MNRAS} \textbf{2007}, \textit{374}, 721
\bibitem[Valtaoja et al.(1989)]{val89} {Valtaoja, L., Valtonen, M.J. \& Byrd, G.G.}, \textit{ApJ} \textbf{1989}, \textit{343}, 47
\bibitem[Valtonen(1996)]{val96b} {Valtonen, M.J.}, \textit{MNRAS} \textbf{1996}, \textit{278}, 186
\bibitem[Valtonen(1996)]{val96}  {Valtonen, M.J.}, \textit{Workshop on Two Years of Intensive Monitoring of OJ~287 and 3C~66A} \textbf{1996}, Proc. meeting at Oxford, England, 11-14 September, 1995. Ed. Leo O. Takalo, U.Turku, 64 
\bibitem[Valtonen \& Karttunen(2006)]{valkar06} {Valtonen, M.J. \& Karttunen, H.} The Three-Body Problem. \textit{Cambridge University Press} \textbf{2006}
\bibitem[Valtonen et al.(2006)]{val06}  {Valtonen, M.J., Lehto, H.J., Sillanp\"a\"a, A., Nilsson, K., Mikkola, S., Hudec, R., Basta, M., Ter\"asranta, H., Haque, S. \& Rampadarath}, \textit{ApJ} \textbf{2006}, \textit{646}, 36-48
\bibitem[Valtonen et al.(2006a)]{val06a}  {Valtonen, M.J., Nilsson, K., Sillanp\"a\"a, A., Takalo, L.O., Lehto, H.J., Keel, W.C., Haque, S., Cornwall, D. \& Mattingly, A.}, \textit{ApJ} \textbf{2006}, \textit{643}, L9-L12
\bibitem[Valtonen(2007)]{val07}  {Valtonen, M.J.}, \textit{ApJ} \textbf{2007}, \textit{659}, 1074-1081
\bibitem[Valtonen et al.(2008)]{val08}  {Valtonen, M.J., Lehto, H.J., Nilsson, K., Heidt,J., Takalo, L.O., Sillanp\"a\"a, A., Villforth, C., Kidger, M., Poyner, G., Pursimo, T. et al.}, \textit{Nature} \textbf{2008}, \textit{452}, 851-853
\bibitem[Valtonen et al.(2010b)]{val10a} {Valtonen, M.~J., Mikkola, S., Lehto, H.J., Hyv\"onen, T., Nilsson, K., Merritt, D., Gopakumar, A., Rampadarath, H., Hudec, R., Basta, M. \& Saunders, R.}, \textit{CeMDA} \textbf{2010}, \textit{106}, 235 
\bibitem[Valtonen \& Sillanp\"a\"a(2011)]{val11b}  {Valtonen, M.J. \& Sillanp\"a\"a}, \textit{AcPol} \textbf{2011}, \textit{51}, 76
\bibitem[Valtonen et al.(2012)]{val12} {Valtonen, M.~J., Ciprini, S. \& Lehto, H.~J.}, \textit{MNRAS} \textbf{2012}, \textit{427}, 77-83
\bibitem[Valtonen et al.(2016)]{val16} {Valtonen, M.~J., Zola, S., Ciprini, S., Gopakumar, A., Matsumoto, K., Sadakane, K., Kidger, M., Gazeas, K., Nilsson, K., Berdyugin, A. et al. (OJ287-15/16 Collaboration).}, \textit{ApJ} \textbf{2016}, \textit{819}, L37
\bibitem[Valtonen et al.(2017)]{val17} {Valtonen, M.~J., Zola, S., Jermak, H., Ciprini, S., Hudec, R., Dey, L., Gopakumar, A., Reichart, D., Caton, D., Gazeas, K. et al.}, \textit{Galaxies} \textbf{2017}, \textit{5}, 83
\bibitem[Valtonen et al.(2019)]{val19} { Valtonen, M. J., Zola, S., Pihajoki, P., Enestam, S., Lehto, H.J., Dey, L.,  Gopakumar, A., Drozd, M., Ogloza, W., Zejmo, M. et al.}, \textit{ApJ} \textbf{2019}, \textit{882}, 88
\bibitem[Valtonen et al.(2021)]{val21} { Valtonen, M. J., Dey, L.; Gopakumar, A., Zola, S.; Komossa, S., Pursimo, T., Gomez, J. L., Hudec, R., Jermak, H. \& Berdyugin, A. V.}, \textit{Galaxies} \textbf{2021}, \textit{10}, 1
\bibitem[Valtonen et al.(2022)]{val22} {Valtonen, M.~J., Zola, S., Ciprini, S., Kidger, M.,Pursimo, T., Gopakumar, A., Matsumoto, K., Sadakane, K., Caton, D.B., Nilsson, K. et al.}, \textit{MNRAS} \textbf{2022}, \textit{514}, 3017
\bibitem[van der Laan(1971)]{van71}{van der Laan, H.}, \textit{Pontificiae Academiae Scientiarum Scripta Varia, Proceedings of a Study Week on Nuclei of Galaxies, ed. D.J.K. O'Connell, Amsterdam: North Holland, and New York: American Elsevier} \textbf{1971}, p. 245
\bibitem[Villata et al.(1998)]{vil98} {Villata, M., Raiteri, C.M, Sillanp\"a\"a, A. \& Takalo, L.O.}, \textit{MNRAS} \textbf{1998}, \textit{293}, L13-L16
\bibitem[Villforth et al.(2010)]{vil10} {Villforth, C., Nilsson, K., Heidt, J., Takalo, L.O., Pursimo, T., Berdyugin, A., Lindfors, E., Pasanen, M. et al.}, \textit{MNRAS} \textbf{2010}, \textit{402}, 2087
\bibitem[Volonteri et al.(2003)]{vol03} {Volonteri, M., Haardt, F. \& Madau, P.}, \textit{ApJ} \textbf{2003}, \textit{582}, 559
\bibitem[Wu et al.(2006)]{wu06} { Wu, J., Zhou, X., Wu, X.-B., Liu, F.-K., Peng, B., Ma, J., Wu, Z., Jiang, Z. \& Chen, J. }, \textit{AJ} \textbf{2006}, \textit{132}, 1256
\bibitem[Zdziarski(1998)]{zdz98} {Zdziarski, A.~A.}, \textit{MNRAS} \textbf{1998}, \textit{296}, L5
\bibitem[Zheng et al.(2008)]{zheng2008} { Zheng, Y.G., Zhang, X., Bi, X.W., Hao, J.M. \& Zhang, H.J. }, \textit{MNRAS} \textbf{2008}, \textit{385}, 823
\bibitem[Zhu \& Thrane(2020)]{Zhu2020} {Zhu, X.-J. \& Thrane, E.}, \textit{ApJ} \textbf{2020}, \textit{900}, 117
\bibitem[Zola et al.(2021)]{zola21} {Zola, S., Kouprianov, V.~V., Reichart, D.~E., Bhatta, G., Caton, D.~B.}, \textit{RMxAC}, \textbf{2021}, \textit{53}, 206
\end{thebibliography}
\end{document}